\documentstyle[aps,eqsecnum,epsfig]{revtex}
\input epsf.sty

\def\laq{\raise 0.4ex\hbox{$<$}\kern -0.8em\lower 0.62ex\hbox{$\sim$}}
\def\gaq{\raise 0.4ex\hbox{$>$}\kern -0.7em\lower 0.62ex\hbox{$\sim$}}
\def\la{\mathrel{\mathpalette\fun <}}
\def\ga{\mathrel{\mathpalette\fun >}}
\def\fun#1#2{\lower3.6pt\vbox{\baselineskip0pt\lineskip.9pt
\ialign{$\mathsurround=0pt#1\hfil##\hfil$\crcr#2\crcr\sim\crcr}}}

\newcommand{\Mpl}{M_{\rm Pl}} 
\newcommand{\mpl}{m_{\rm Pl}}
\newcommand{\Ms}{M_{\rm S}}
\newcommand{\GeV}{{\rm GeV}}

\newcommand{\pa}{\partial} 
\newcommand{\vphi}{\varphi} 
\newcommand{\beq}{\begin{equation}} 
\newcommand{\eeq}{\end{equation}}
\newcommand{\bea}{\begin{eqnarray}} 
\newcommand{\eea}{\end{eqnarray}}

\begin{document}

\title{\Large\bf Reheating and dangerous relics \\
in pre-big bang string cosmology}

\author{Alessandra Buonanno${}^{a}$, Martin Lemoine${}^{b}$ and Keith A. Olive${}^{c}$}

\address{$^a$ {\it Theoretical Astrophysics and Relativity Group,\\
California Institute of Technology, Pasadena, CA 91125, USA }\\ {$^b$
{\it DARC, UMR--8629 CNRS, Observatoire de Paris-Meudon, F-92195
Meudon, France}} \\ {$^c$ {\it Theoretical Physics Institute, School of
Physics and Astronomy,\\ University of Minnesota  Minneapolis, MN 55455,
USA}}}


\maketitle
\begin{abstract}
We discuss the mechanism of reheating in pre-big bang string cosmology
and we calculate the amount of moduli and gravitinos produced
gravitationally and in scattering processes of the thermal bath. We
find that this abundance always exceeds the limits imposed by big-bang
nucleosynthesis, and significant entropy production is required.  The
exact amount of entropy needed depends on the details of the high
curvature phase between the dilaton-driven inflationary era and the
radiation era. We show that the domination and decay of the zero-mode
of a modulus field, which could well be the dilaton, or of axions,
suffices to dilute moduli and gravitinos. In this context,
baryogenesis can be accomodated in a simple way via the Affleck-Dine
mechanism and in some cases the Affleck-Dine condensate could provide
both the source of entropy and the baryon asymmetry.
\end{abstract}

\vskip -4.5in \rightline{UMN--TH--1906/00} \rightline{TPI--MINN--00/26}
\rightline{GRP/00/541} \rightline{hep-th/0006054}
\vskip 4.0in 

\section{Introduction}
\label{sec1}

In the pre-big bang scenario (PBB) \cite{PBB}, the standard
Friedman-Robertson-Walker (FRW) post-big bang picture emerges as the
late-time history of a Universe which, in a prehistoric era (the so
called pre-big bang era), underwent an inflationary expansion driven
by the growth of the universal coupling of the theory.  This latter
phase is also referred to as dilaton driven inflation (DDI) as its
super-inflationary dynamics are driven by the kinetic energy of the
dilaton field.  A crucial difference between the pre-big bang model
and standard inflationary theories, is that in the PBB, the universe
starts its evolution in a classical state, the most general
perturbative solution of the tree-level low-energy string effective
action.  The analysis of this initial state, and its naturalness, has
led to a debate, as to whether the initial conditions needed to solve
the horizon and flatness problems can be deemed natural~\cite{IC}. The
problem of the graceful exit of the inflationary era is also an
unsolved question. No-go theorems preventing the branch change from
DDI to the dual solution of the FRW type have been demonstrated when
either an axion field and a dilaton/axion potential or stringy fluid
sources have been introduced in the tree-level effective
action~\cite{KMO95}.  It is now recognized that if the branch change
from DDI to FRW expansion is to occur, it should arise as a
consequence of quantum loops effects and/or high curvature
corrections, and encouraging progress has been made in this direction
\cite{GMV97},\cite{BM97etal}. Nevertheless, the PBB model remains an
attractive variant to standard inflationary cosmology, notably since
the initial state of the Universe lies in the weakly coupled regime of
string theory, and its dynamics are thus well controled by the
tree-level low-energy effective action. This is in contrast with
standard inflationary theories which generically experience
difficulties in extracting a well-suited Lagrangian for the inflaton
field from a well-defined underlying fundamental theory, and in which
the initial inflaton field values are usually of order the Planck
scale.

In recent years, significant effort has been spent on understanding
the physics of the pre-big bang initial state and its high curvature
phase, and on extracting observational predictions for this
scenario. In this respect, one should note the prediction of a
stochastic background of gravitational waves~\cite{GW}, whose
amplitude might be well above that predicted by models of standard
inflation, as well as the amplification of quantum vacuum
electromagnetic fluctuations~ \cite{EW}, due to the non-conformal
coupling between the gravitational and the electromagnetic fields.
More recently, it has been shown that the amplification of (universal)
axion quantum fluctuations might provide adequate seeds for the
formation of large-scale structures, and the resulting large and small
angular scale anisotropies of the cosmic microwave background have
been calculated \cite{DGSV98,MVDV99}. This provides a characteristic
signal of non-Gaussian isocurvature perturbations\footnote{One should
note here that the recent high precision small angular scale data of
the BOOMERANG and MAXIMA experiments~\cite{BM} do not seem to confirm
the predictions made in Refs.~\cite{DGSV98,MVDV99}.}. So far little
attention has been paid to the phenomenology of the post-big bang FRW
era, and notably on the mechanism of reheating. It has been proposed
that reheating could proceed via gravitational particle production
\cite{V94}. However, as we argue here, this predicts the presence of
too many dangerous relics, much like non-oscillatory inflationary
models \cite{FKL99}, and notably an abundance of gravitationally
interacting scalars (e.g. moduli) well in excess of the limits imposed
by big bang nucleosynthesis (BBN) on the abundance of late decaying
massive particles. On top of that one naively expects that in some PBB
scenarios the FRW era starts at a high Hubble scale, of order the
string scale $\sim 10^{17}\,{\rm GeV}$, and therefore gravitinos and
moduli should also be produced copiously in scattering processes in
the thermal bath.

These considerations warrant the present detailed study of the physics
and phenomelogical problems of the PBB scenario in the post-pre-big
bang era. We will find that indeed, for the variants of the PBB
scenario hitherto proposed, there is inevitably need for significant
entropy production. However, as we will argue, there exist various
possible and natural sources of entropy production in pre-big bang models,
notably the domination and decay of the zero-mode of a modulus, of the
dilaton, or of axions, depending on the masses of these fields, which
can dilute effectively moduli, gravitinos and monopoles. We will also
show that this allows one to efficiently implement  Affleck-Dine
baryogenesis. In the following, we will thus focus on reheating, on
the gravitino/moduli problem, and on the origin of the baryon
asymmetry of the Universe. We will try to remain as general as
possible, in particular with respect to the possible presence of an
intermediate phase between DDI and FRW.

The outline of this paper is as follows. In Section II we review the
dynamics of the PBB model, and some variants (as far as the
intermediate phase is concerned) proposed in the literature. In
Section III, we provide a book-keeping of the particle content of the
Universe at the beginning of the FRW era, and explicitly show the need
for entropy production. In Section IV we review the various
possibilities for entropy production in the context of the PBB model
when the transition from DDI to FRW occurs suddenly, and discuss
baryogenesis. We defer the study of intermediate phases to Sections V
and VI, since the consequences in that case are different but the
logic of the argument is the same. We summarize our results in Section
VII.

\section{The pre-big bang era}
\label{sec2}

Let us first start by reviewing the different eras and dynamics
envisaged in the PBB model.  We shall restrict ourselves to the
four-dimensional tree-level low-energy string effective action derived
from heterotic string theory compactified on a six torus
\cite{MS93,BMUV98}, whose bosonic sector is described by:~\footnote{We use the
conventions $(-,+,+,+)$ and ${\cal R}^\mu_{\,\,\,\,\nu \rho \sigma} =
\Gamma^\mu_{\nu\sigma,\rho} - \dots\,,\,\,{\cal R}_{\mu \nu} = {\cal
R}^\rho_{\,\,\,\,\mu \rho \nu}$.  Units are $\hbar=k=c=1$,
$\mpl=\Mpl/(8\pi)^{1/2}\simeq2.4\times10^{18}\,$GeV is the reduced
Planck mass}

\beq
\label{A1}
S_{\rm eff} = \frac{1}{2 \lambda_s^2}\, \int d^4 x
\sqrt{-g}\,e^{-\vphi}\,\left [ {\cal R} + g^{\mu \nu}\,\pa_\mu
\vphi\,\pa_\nu \vphi - g^{\mu \nu}\,\pa_\mu \sigma_a\,\pa_\nu \sigma_a
+ {\cal L}_{\rm matter} \right ]\,, \eeq 
where $\lambda_s = \sqrt{\alpha^\prime} = \sqrt{8 \pi}/M_s$ is the
string-length parameter, $M_s$ denotes the string mass and the
following relations hold: 

\beq \frac{e^{-\varphi}}{\lambda_s^2} = \frac{1}{l_{\rm pl}^2} =
\frac{1}{8 \pi\,G_N}\,, \quad \quad \alpha_{\rm GUT}(\lambda_s^{-1}) =
\frac{g^2}{4\pi}\,, \quad \quad g^2 = e^\vphi\,.  \eeq 

Henceforth, we shall assume that at the present time $g_0 \sim 0.01
- 0.1$.  We restrict the internal compact
space to a diagonal metric  $g_{a b} = e^{2 \sigma_a}\,\delta_{a
b}$ with $a,b = 4, \dots, 9$, and we denote by $\vphi$ the effective
four-dimensional dilaton field $\vphi = \phi_{10} - \sum_a \sigma_a$. The
matter Lagrangian,
${\cal L}_{\rm matter}$, is composed of scalars, gauge fields and
axions. We assume that the gauge and axion fields do not contribute to the
cosmological background, i.e. we deal only with their quantum vacuum
fluctuations. For the gauge fields, we consider the heterotic
gauge field $A_\mu$ and the Kaluza-Klein gauge fields related to
internal components of the metric and the three-form $H_{\mu\nu\rho}$,
respectively $V_\mu^{\,\,a}$ and $W_\mu^{\,\,a}$~\cite{MS93,BMUV98}:

\beq
\label{A4}
{\cal L}_{\rm gauge\, fields} = \frac{1}{4}\,\alpha^\prime\, F_{\mu
\nu}\,F^{\mu \nu} - \frac{1}{4}\,\alpha^\prime\,e^{2\sigma_a}\,V_{\mu
\nu}^{\,\,\,\,a}\,V^{\mu \nu a} - \frac{1}{4}\, \alpha^\prime\, e^{-2
\sigma_a}\,W_{\mu \nu a}\,W^{\mu \nu }_{\,\,\,\,\,a}\,, \eeq
where  

\beq
\label{A5}
F_{\mu \nu } = \partial_\mu A_{\nu} -\partial_\nu A_{\mu} \,, \quad
\quad V_{\mu \nu}^{\,\,\,\,a} = \partial_\mu V_\nu^{\,\,a}
-\partial_\nu V_\mu^{\,\,a} \,, \quad \quad W_{\mu \nu a} =
\partial_\mu W_{\nu a} -\partial_\nu W_{\mu a} \,.  \eeq

Finally, for the axion fields we have

\beq
\label{A6}
{\cal L}_{\rm axions} = -\frac{1}{2}\,e^{2 \varphi}\,g^{\mu
\nu}\,\pa_{\mu}A\,\pa_\nu A -\frac{1}{4}\,g^{\mu
\nu}\,e^{-2\sigma_b}\, e^{-2\sigma_c}\,\partial_\mu B_{b
c}\,\partial_\nu B_{b c}\,, \eeq
where $B_{a b}$ is the pseudo-scalar field associated to the
compactified components of the anti-symmetric field living in ten
dimensions, while $A$ is the axion related to the anti-symmetric
tensor $H^{\mu \nu \rho}$ in four dimensions by the usual relation
$H^{\mu \nu \rho} \equiv \epsilon^{\mu \nu \rho \sigma}\,e^\vphi\,\pa_\sigma
A$ (where $\epsilon^{\mu \nu \rho \sigma}$ is the covariant full
antisymmetric Levi-Civita tensor).

Let us consider first the simplest scenario in which the universe
undergoes a super-inflationary evolution up to conformal time
$\eta_1$, at Hubble scale $H_1 \sim M_s$ and where the radiation
dominated era is supposed to start, i.e. where the branch change from
DDI to FRW occurs.  The cosmological background during such DDI era is
given, in conformal time, by

\bea
\label{A7}
&& a(\eta) \sim (-\eta)^{\delta/(1-\delta)} \,, \quad \quad \vphi \sim
\vphi_1 + \frac{3 \delta -1}{1-\delta}\,\log (-\eta)\,, \\ && \sigma_a
\sim \sigma_{a 1} + \frac{\beta_a}{1-\delta}\,\log(-\eta) \,, \quad
\quad \delta < 0\,, \quad \quad \beta_a > 0\,, \eea 
(note that with respect to the cosmological time $a(t) \sim
(-t)^\delta$).  
A scalar field $\chi$ with canonical kinetic term evolves classically
during DDI as:

\beq
\label{A8}
\chi =  \frac{1}{\sqrt{\alpha^\prime}}\,
\frac{\beta_\chi}{1-\delta}\,\log(-\eta) \,, \eeq where the
parameters $\delta$, $\beta_a$, and $\beta_\chi$ satisfy the
Kasner-type constraint:

\beq
\label{A9}
1 = 3\,\delta^2 + \sum_a \beta_a^2 + \frac{1}{2}\, \beta_\chi^2\,.
\eeq This Kasner constraint can be rewritten as a relation between
$\delta$ and an effective set of $\beta_a$ which parametrizes the
evolution of other scalars, including internal moduli. We will thus
neglect $\beta_\chi$ in the following.  We have also assumed that the
four-dimensional non-compact space-time expands isotropically, while
the contraction of the six internal dimensions can be
anisotropic. After the branch change has occurred, the metric is that
of a spatially flat FRW space-time; at that point the kinetic energy
of the dilaton has become negligible, and the dynamics are thus driven
by radiation, so that $a(\eta>\eta_1)\propto \eta$. The ulterior
evolution of the dilaton is an unsolved question. We will assume that
the dilaton is fixed in the radiation era \cite{tsey}, but we will also
indicate explicitly the dependence on the string coupling $g_1$
(corresponding to the value of the coupling at the start of the radiation
dominated phase) in our results. In particular, in the radiation era, the
critical energy density $\rho_c$ as a function of the Hubble scale $H$
is: $\rho_c=(3/8\pi)g_1^{-2}H^2M_s^2$.

In some pre-big bang scenarios, the branch change from DDI to FRW is
not instantaneous, and one considers an intermediate phase whose
dynamics are obtained by taking into account higher order corrections
to the low-energy effective action, such as finite size string effects
and quantum string-loop effects.  Unfortunately, a thorough knowledge
of the dynamics and duration of this intermediate phase is still
lacking. Cosmological solutions, which partially describe the high
curvature phase, have nevertheless been proposed in the literature,
most notably: (i) the ``string'' intermediate era, obtained by solving
the equations of motion with only the first order corrections in
$\alpha^\prime$ included \cite{GMV97}, and (ii) the ``dual-dilaton''
intermediate phase~\cite{BMUV98} (see also \cite{MG96} where 
this scenario was discussed in the more general framework 
of non-minimal models ), where one assumes that all $\alpha'$
corrections are sufficient to provide by themselves (without including
string-loop effects) a sudden branch-change from the DDI to another
duality-related vacuum phase of the FRW type.
In the ``string'' intermediate phase, the Hubble parameter is constant 
hence the dynamics
is inflationary in the string frame, while in the ``dual-dilaton''
era, the Hubble parameter makes a bounce around its maximal value at
the string mass. To simplify the discussion, we shall often assume in
both cases that the internal dimensions have been stabilized in some
way before the Universe enters the intermediate era.  We fix at
$\eta_s$ the time and $H_s$ the Hubble scale at which the Universe
transits from the DDI era to the intermediate phase.

For the ``string'' intermediate phase, one obtains \cite{GMV97}: 

\beq
\label{A10}
a(\eta) \sim - \frac{1}{H_s\,\eta}\,, \quad \quad \vphi(\eta) \sim
\vphi_s -2 \zeta\,\log(-\eta)\,, \quad \quad \zeta \geq 0\,, \eeq
where $\zeta$ is an arbitrary parameter which governs the growing of
the dilaton field, while with the ``dual-dilaton'' era~\cite{BMUV98}

\beq a(\eta) \sim (-\eta)^{\theta/(1-\theta)} \,, \quad \quad \vphi
\sim \vphi_s + \frac{3 \theta -1}{1-\theta}\,\log (-\eta)\,, \eeq

and $\theta$ satisfies a Kasner constraint similar to Eq.~(\ref{A9}).

\section{The post--pre-big bang era}
\label{sec3}

In this section, we will consider the simplest version of the
pre-big bang scenario with a sudden branch change from DDI to FRW,
i.e. no intermediate phase of dynamics. 

\subsection{Particle content due to gravitational production}
\label{subsec3.1}

The particles present at the very beginning of the radiation era
result from gravitational particle production, in contrast to standard
inflationary models, in which the post-inflationary era is dominated
by inflaton condensates, which later decay into radiation in the
reheating process. In fact, in the PBB scenario it is the kinetic
energy of the dilaton, which drives the DDI phase, that is converted
into gravitationally created particles, whose energy density will
drive the FRW era.  One can provide a simple estimate of the energy
density contained in fields subject to gravitational particle
creation, when no intermediate phase is present.  If $a_1\equiv
a(\eta_1)$ is the scale factor at the branch change, then $\eta_1=
(a_1H_1)^{-1}$ (since $H\equiv a'/a^2$), and $k_1=1/\eta_1$ represents
a comoving wavenumber corresponding to the horizon size at the branch
change. We also define $d\rho_j/d\ln k$ as the energy density spectrum
in particle species $j$ as a function of wavenumber $k$. Then one
obtains $d\rho_j/d\ln k \simeq 0$ for wavenumbers $k>k_1$, since those
modes have remained within the horizon at all times, and could not be
excited by the gravitational field. For fluctuations that exited the
horizon during DDI and re-entered during FRW, i.e. those modes with
wavenumber $k<k_1$, one generically obtains $d\rho_j/d\ln k \propto
(k/k_1)^{n_j^{\rm \tiny DDI}}$, and ${n_j^{\rm \tiny DDI}}$ is the
spectral index acquired by species $j$ due to the dynamics of the DDI
phase and transition into FRW. One imposes ${n_j^{\rm \tiny DDI}}>0$
so as to avoid infrared divergences, i.e. large-scale inhomogeneities
(see also below) and the energy density in species $j$ is dominated by
the energy density in the log interval around $k_1$, so that
$\rho_j\sim (d\rho_j/d\ln k)_{|_{k_1}}$. Moreover, $k_1$ corresponds
to the maximal amplified wavenumber: this mode has exited and
re-entered the horizon at the same time, and roughly one particle has
been produced in that mode. Gravitational particle production thus
respects a democracy rule \cite{V94}, namely all species share roughly
the same energy density $\sim (a_1/a)^4 H_1^4$, corresponding to one
particle produced with momentum $H_1$ in phase space volume $\sim
H_1^3$. Therefore, for all species $j$, $\rho_j\sim (a_1/a)^4 H_1^4$
at times $\eta>\eta_1$, and consequently $\Omega_j \sim g_1^2
(H_1/M_s)^2$, where $\Omega_j$ denotes the density parameter in
species $j$ and $g_1 = e^{\vphi_1/2}$ is the value of the string
coupling at the beginning of the radiation era [$\Mpl(\eta_1) =
M_s/g_1$].

When more accurate calculations are performed, one finds that the
above democracy rule is satisfied to within less than an order of
magnitude between different species, and one obtains (with ${n_j^{\rm
\tiny DDI}} \neq 0$) 

\beq
\label{A11}
\frac{d \rho_j}{d \log k}(\eta) \simeq \frac{{\cal N}_j}{2 \pi^2}\,
\left(\frac{a_1}{a}\right)^{4}H_1^4\,\left ( \frac{k}{k_1} \right
)^{n_j^{\rm \tiny DDI}}\quad\quad \eta>\eta_1\,, \eeq and 

\beq
\label{A12}
\Omega_j(\eta) \simeq \frac{4\, {\cal N}_j}{3 \pi}\,
\left(\frac{H_1}{M_s} \right )^2 \,g_1^2\, \left | 1 - \left
(\frac{k_{\rm H}}{k_1} \right )^{n_j^{\rm \tiny DDI}} \right |
\quad\quad \eta>\eta_1\,, \eeq where ${\cal N}_j$ is the number of
helicity states in species $j$.  In Eq.~(\ref{A12}), $k_{\rm H}$ is
the wavenumber corresponding to the horizon size at time $\eta$,
i.e. $k_{\rm H}\equiv1/\eta$. In effect, only modes whose wavelength
is smaller than the horizon size can be thought as propagating as
particles, and can be included in the energy density. For modes whose
wavelength is larger than the horizon size, the definition of an
energy density becomes gauge-dependent. Nevertheless, since we impose
$n_j^{\rm \tiny DDI}>0$ to avoid infra-red divergence problems, the
contribution from the term in the absolute value in Eq.~(\ref{A12}) is
negligible for $\eta\gg\eta_1$, and the density parameter reduces to
that deduced above by heuristic arguments, up to the fudge factor
$4{\cal N}_j/3\pi$. At this point, one should note that some fields
$j$, and notably the axion $A$, can actually have $n_j^{\rm \tiny
DDI}<0$~\footnote{The fact that axion fields can have negative
spectral slopes is not a prerogative of the heterotic string model
under study, in fact Copeland et al. \cite{CLW98} have shown that in
the type IIB string model, with three axion fields, one of them at
least must have $n_j^{\rm \tiny DDI} <0$, which can pose serious
problems for the PBB model. On the other hand it has been shown
recently \cite{BW00} that with a $SL(4,\Re)$-invariant effective
action, there exists a region of parameter space where all the axions
have $n_j^{\rm \tiny DDI} >0$.} (see Tab.~\ref{tab1}).  Again, to
avoid infra-red problems, we shall impose $n_A^{\rm \tiny DDI} > 0$,
that is $-1/3 < \delta < 0$~\cite{CEW97}.  For the particular case of
the PBB dynamics, it has been shown that the fields subject to
particle production are those of spin 0, 1 and 2. One should mention
that in Einstein gravity, abelian gauge fields are conformally
invariant, and thus not gravitationally amplified; here, their
conformal invariance is broken by the time evolution of the string
coupling. Fermions (spin 1/2 and 3/2) are not produced \cite{BH00}
(see also Refs.~\cite{KKLVP99,GRT99a,ML99,GRT99b}), at least when effects
of compactification are neglected (see below).

\begin{table}
\begin{center}
\begin{tabular}{cccc}
{\sl Particles} & $n^{\rm \tiny \tiny DDI}$ & $n$
{\it \small (string phase)} 
& $n$ {\it \small (dual-dilaton phase)}\\\hline
{\sl moduli} & $3$ &  $ \left\{\begin{array}{ll}
6-2\zeta &  \zeta > \frac{3}{2}\\
2\zeta &  \zeta < \frac{3}{2}
\end{array} \right .$ & $4$\\
{\sl axion A} & $-0.46 \div 1$ & $-2 \zeta$ & $-2.9 \div -1.4$\\
{\sl axion $B_{a b}$} & $- 1 \div 3$&  $4-2\zeta $ & $0 \div 4$\\
{\sl Heterotic photons} & $1 \div 3$ & $4-2\zeta$ & $0.54 \div 1.28$\\
{\sl KK photons} & $1 \div 3$ & $4-2\zeta$ & $ -0.73 \div 0.54$
\end{tabular}
\end{center}
\caption{\sl Spectral slopes for the particles which have been
amplified gravitationally during the PBB phase assuming 
non-dynamical internal dimensions during the intermediate phases.  
The spectral indices in the three columns refer 
to fluctuations that exited the horizon
during DDI, during the ``string'' phase, 
or re-entered during the ``dual-dilaton'' phase, respectively.}
\label{tab1}
\end{table}

In Tab.~\ref{tab1} we summarize the values of the spectral slopes for
all of the particles present in the model \cite{BMUV98} assuming, for
simplicity, non-dynamical internal dimensions during the intermediate
phases.  The range of values of $n^{\rm \tiny \tiny DDI}$ have been
obtained varying $\delta$ and considering the possibility of having
either one or six internal dynamical dimensions during DDI. Note that
the spectral slope $n^{\rm \tiny \tiny DDI}$ for the moduli fields
have been obtained while considering them as part of the
background~\cite{CEW97}, while in the determination of $n$ with an
intermediate phase, we neglect their presence in the
background. Spectral slopes for other scalars depend in principle on
their kinetic terms, and for simplicity we will assume that they have
the same slopes as moduli fields, i.e. that they have canonical
kinetic terms. The spectral slopes $n$ in the second and third columns
correspond to the slopes for fluctuations that exited during the
``string'' phase and re-entered during FRW, or exited during DDI and
re-entered during the ``dual-dilaton'' era, respectively, and will be
discussed in Sections VI and V.

\subsection{Thermalization and reheating}
\label{subsec3.2}

In this section we analyse the thermalization and reheating process
due to gravitational particle production. Let us start by considering
the simple generic case with no intermediate phase. At Hubble scale
$H_1$, all fields are produced with similar energy density
(``democracy rule''). Let us denote by ${\cal N}_r$ the number of
degrees of freedom in spin 0 and spin 1 fields charged under the gauge
groups of the observable sector. Similarly, if ${\cal N}_{\rm tot}$
denotes the total number of degrees of freedom in spin 0, 1 and 2,
{\it i.e.} that of the fields produced gravitationally, then the
democracy rule implies that the fraction of energy density contained
in radiation (in the observable sector) is $\Omega_r\simeq {\cal
N}_r/{\cal N}_{\rm tot}$. If the number of particles charged under
gauge group is much larger than the number of gauge singlets, we get
$\Omega_r \sim 1$. However, in some string models, the number of gauge
singlets may actually exceed the number of charged states, and in this
case, one would generically expect $0.01\la\Omega_r <1$.  Henceforth,
to keep the discussion generic we shall explicit the dependence on
$\Omega_r$.

All fields carry typical energy $E\sim H_1 (a_1/a)$, and the radiation
number density $n_r\sim \Omega_r\rho_c/E$. Gauge non-singlets interact
with cross-section $\sigma\sim \alpha^2/E^2$ and thus, thermalization
occurs when the interaction rate $\Gamma_r\equiv n_r\sigma v\ga H$
($v\sim1$ is the relative velocity), i.e. at scale factor $a_{\rm
th}$:

\beq
\label{A21}
\frac{a_{\rm th}}{a_1} \simeq \,{\rm max}\left[1\,,\,8 \Omega_r^{-1}
g_1^2\alpha^{-2} \left(\frac{H_1}{M_s}\right)^2\right]\,.  
\eeq 

For $\Omega_r\sim 1$, $g_1\sim 10^{-1}$, $\alpha = g_1^2/(4\pi) \sim
10^{-3}$ and $H_1\sim M_s$, thermalization occurs in $\simeq9$
e-foldings of the scale factor at $H_{\rm th} \sim 10^{-8}\,H_1$.  Let
us observe that the value of $\alpha$ we have used differs from
$\alpha_{\rm GUT} \sim 1/20$; this discrepancy, which is linked to the
difference between the string scale $M_s$ and the GUT scale, is
usually attributed to threshold effects. Note also that the various
gauge fields $A_\mu$, $V_\mu^a$ and $W_\mu^a$ we introduced in the
action, Eq.~(\ref{A4}), can in principle have different gauge
couplings depending on the compactification. For simplicity we assume
a single coupling constant, $\alpha=g_1^2/4\pi$, which refers to the
$A_\mu$ field.

Even before thermalization is achieved, one can define an effective
entropy density $s=(4/3)\rho_r/T_r$, where
$T_r=(30/\pi^2)^{1/4}g_{\star}^{-1/4}\rho_r^{1/4}$ is an effective
temperature, with $g_\star$ the number of degrees of freedom in the
radiation after thermalization, i.e. including spin 1/2 fields that
were not produced gravitationally but were re-created in scattering
processes. This effective entropy will reduce to the standard entropy
of the radiation once thermalization has been achieved, and $s \simeq
0.2g_\star^{1/4}\Omega_r^{3/4}(H_1 M_s/g_1)^{3/2}(a_1/a)^{3}$. Then,
if $\Omega_r\simeq1$, reheating is complete once radiation has
thermalized. If $\Omega_r < 1$, reheating would only be achieved once
the fields that carry the remainder of the energy density have decayed to
radiation. Such processes are constrained by big bang nucleosynthesis,
which requires that at temperatures $T_r \la 1\,$MeV, $\Omega_r\simeq
1$ to within a few percent. Nevertheless, as we will argue in the
following subsections, it will be necessary to release a vast amount
of entropy to dilute the dangerous relics produced. This entropy
production may be viewed as a period of secondary reheating.

\subsection{Dangerous relics}
\label{subsec3.3}

Using the above results, one can determine the number density of
scalar fields with gravitational interactions present at the beginning
of the radiation era and analyse their possible phenomenological
consequences on BBN.  In what follows, we will denote such scalar
fields generically as moduli. Moduli are produced gravitationally as
argued above, and one also expects them to be produced in scatterings
of the thermal bath at time $\eta>\eta_1$. We will inspect each of
these effects in turn, and discuss moduli and gravitinos.

\subsubsection{Moduli}
\label{subsubsec3.3.1}

We adopt the generic notation $Y_j=n_j/s$ for the number-density $n_j$
to entropy-density $s$ ratio of species $j$; the entropy density in
radiation can be written as before $s\simeq
g_\star^{1/4}\rho_r^{3/4}$. Using Eq.~(\ref{A11}) for $\rho_r$, one
can rewrite $s\sim (2\pi^2)^{-3/4}g_\star^{1/4}{\cal N}_r^{3/4} H_1^3
(a_1/a)^{3}$. Note the dependence on ${\cal N}_r^{3/4}$ which counts
the number of degrees of freedom produced gravitationally, namely
those of spin 0 and 1. Due to supersymmetry, obviously ${\cal N}_r\sim
g_\star/2$, since $g_{\star}$ accounts for these latter and their
supersymmetric partners.

Similarly, the number density of moduli $n_m\simeq \rho_m/E_m$, with
typical energy $E_m \simeq H_1 (a_1/a)$, hence from Eq.~(\ref{A11})
$n_m\simeq (1/2\pi^2) H_1^3 (a_1/a)^{3}$, and:

\beq
\label{Y_mg}
Y_m^g \simeq 0.3 g_\star^{-1}\,,
\eeq
where the superscript $g$ on $Y_m$ refers to gravitational
production. 
For $g_\star\sim 10^2-10^3$, Eq.~(\ref{Y_mg})
gives 
$Y_m^g\sim10^{-4}-10^{-3}$, which is well above the bounds imposed by
BBN on the abundance of late decaying massive particles see e.g.,
\cite{wein}.  Indeed, let us contrast these estimates with the upper
limits on $Y_m$ imposed by big bang nucleosynthesis (BBN) from photon
injection. When applied to the case of moduli and gravitinos whose
lifetime $\sim \gamma^{-1}\Mpl^2/m^3$, where $m\sim {\cal O}(100\,{\rm
GeV})$ denotes the modulus/gravitino mass, and $\gamma$ is a fudge
factor for the decay width ($\gamma\simeq1/4$ in the case of the
gravitino) these constraints become \cite{EKN84,HKKM99}:
$Y_m\la10^{-15}$ for $m\simeq100\,$GeV, $Y_m\la10^{-14}$ for
$m\simeq300\,$GeV, and $Y_m\la5\times10^{-13}$ for
$m\simeq1\,$TeV~\footnote{Note that Holtmann et al. define $Y_m$ with
respect to the photon number density $n_\gamma$, not $s$, and today
$s\simeq7.0n_\gamma$; also, the constraints quoted assume $\gamma=1$;
for $\gamma\neq1$, the mass estimates apply to $\gamma^{1/3}m$,
instead of $m$.}. These bounds assume that the gravitino/modulus
decays into photons with a branching ratio unity. Results weaker by
$\sim1-2$ orders of magnitude would be obtained if the
modulus/gravitino decays only into neutrinos, since high energy
neutrinos produce an electromagnetic shower by interacting with the
cosmic neutrino background~\cite{KM94}. Moreover, stringent
constraints in the high mass range $m\ga 1\,$TeV would also be
obtained if hadronic decay is allowed~\cite{S96}. Thus a safe and
generic limit is $Y_m \la 10^{-13}$, which corresponds to the
celebrated limit on the reheating temperature $T_{\rm RH}\la10^9\,$GeV
in standard inflationary scenarios. When considering these limits and
the above results for the PBB scenario, one realizes that entropy
production to the level of at least $\sim 8 - 10$ orders of magnitude is
required.

Moduli are also created in scattering processes of the thermal
bath. The 
total amount of moduli present, at times $\eta\gg\eta_1$,
can be obtained by solving the Boltzmann equation with adequate
production and destruction terms, with $Y_m=Y^g_m$ as initial
condition at $\eta=\eta_1$. This equation, when written as a function
of radiation temperature $T$ reads:

\begin{equation}
\frac{dY^s_m}{dT} = -\left [ \frac{1}{2}\,\sum_{a,b}\,Y_a\,Y_b \sigma_{a+b\to
m+c} - \sum_{c}\,Y_m\,Y_c\,\sigma_{c+m\to a+b}\right ]\,\frac{s}{HT}\,,
\label{Y_B}
\end{equation}
where the superscript $s$ on $Y_m$ refers to moduli produced by
scattering processes. In the above equation, $Y_{a,b,c}$ denotes the
number density to entropy density ratio of species $a,b,c$, and
$\sigma_{a+b\to m+c}$ is the cross-section of the process $a+b\to
m+c$. Generically, $a$ and $b$ are relativistic, in which case
$Y_{a,b}\simeq0.3/g_\star$ [note that the first sum in Eq.~(\ref{Y_B})
is over degrees of freedom of $a$ and $b$].  Since we found
previously $Y^g_m\sim0.3/g_\star\sim Y_{a,b,c}$, which corresponds to
equilibrium with radiation, the Boltzmann equation implies
$dY_m^s/dT\sim 0$, i.e. the production/destruction of moduli in
scattering processes is negligible as compared to $Y^g_m$, and the
final $Y_m\sim Y^g_m$.

\subsubsection{Gravitinos}
\label{subsubsec3.3.3}

It has been argued recently~\cite{BH00} that gravitinos should not be
produced gravitationally in the PBB scenario, if the gravitino is
effectively massless, {\it i.e.} if the superpotential $\langle W
\rangle\sim0$ in the DDI and FRW eras. During the DDI phase, one
indeed expects $\langle W \rangle=0$ in a simple model.  However,
compactification of internal dimensions during DDI or non-perturbative
effects to stabilize the dilaton in FRW should lead to the appearance
of a superpotential, which would break the above condition, and result
in gravitino production. Unfortunately, the magnitude of this mass
term is very model-dependent and one cannot really determine the
amount of gravitinos produced gravitationally. However, it should be
noted that if one gravitino is produced per mode around the branch
change frequency, corresponding to saturation of Fermi-Dirac
statistics, one would find $Y^g_{3/2}\sim 0.3/g_\star$ per helicity
state as in the case of moduli.

In any case, gravitinos are produced in scatterings of the thermal
bath, in the same fashion as moduli and the Boltzmann equation
(\ref{Y_B}) can be used substituting $m \to 3/2$ etc. If
$Y^g_{3/2}\sim 0.3/g_\star$ per helicity state, corresponding to
equilibrium, then as before $dY_{3/2}/dT\sim 0$. However if, as
advocated in Ref.~\cite{BW00}, gravitinos are not produced
gravitationally in the PBB scenario, then $Y^g_{3/2}=0$, and an
estimate of $Y_{3/2}$ is given by integrating the Boltzmann equation,
neglecting annihilation and co-annihilation channels. This neglect is
justified as long as the final value $Y_{3/2}\ll 1/g_\star$, i.e. as
long as equilibrium is not reached. Thus one obtains, using the
Boltzmann equation for gravitinos, the simple result
\cite{eln,nos,ehnos} \beq Y_{3/2}^s \simeq {\Gamma \over H}\,Y_a
\label{Y_32b}
\eeq
evaluated at the end of the PBB phase, where $\Gamma$ is the gravitino 
production rate and $Y_a$ denotes the number density to entropy 
density ratio in species a. For the particle content of the MSSM,
the gravitino total production cross-section is $\Sigma_{\rm tot}\simeq
250\alpha/\mpl^2$~\cite{M95,EKN84} (see also \cite{enor} for
finite-temperature contribution to the gravitino production
cross-section), and therefore, integration of the Boltzmann equation
(assuming radiation domination $a\propto \eta$) gives:

\begin{equation}
Y^s_{3/2} \simeq 100 \alpha\,g_\star^{-7/4}\, g_1^{1/2}\,\Omega_r^{3/4}\,
\left (\frac{H_1}{M_s} \right )^{1/2}.
\label{Y_32s}
\end{equation}

For $g_1 \sim 10^{-1}$, $\alpha = g_1^2/(4 \pi) \sim 10^{-3}$,
$\Omega_r\simeq 1$, $g_\star\sim 10^2-10^3$ and $H_1\sim M_s$, one
thus finds $Y^s_{3/2}\sim 10^{-7}-10^{-5}$; this justifies our neglect
of the annihilation channels in the Boltzmann equation. In any case,
the destruction terms would ensure that $Y_{3/2}$ would never exceed
its equilibrium value, so that $Y_{3/2}\sim {\rm min}\left[Y^s_{3/2} +
Y^g_{3/2}, 1/g_\star\right]$ is a good approximation to the final
abundance of gravitinos, independently of the value of $Y^g_{3/2}$.

  It should be noted that Eq.~(\ref{Y_32s}) evaluates the number of
gravitinos produced before radiation has thermalized [see
Eq.~(\ref{A21})], at the Hubble scale $H_1$. Moreover, if fermions are not
produced gravitationally, then the only charged non-singlets present
at scale $H_1$ are those of spin 0 and 1, and the gravitino production
cross-section should be smaller, since only channels
$a+b\to\Psi_{3/2}+c$ involving $a,b$ of spin 0 or 1 should
contribute. However, we do not expect this uncertainty to exceed an
order of magnitude~\cite{eln,nos,ehnos}. Furthermore, we used
$\alpha\sim10^{-3}$ as before, corresponding to $g_1\sim0.1$, and
we neglected the running of $\alpha$ between $M_s$ and $M_{\rm
GUT}$. However, it is easy to check that calculating $Y_{3/2}$ with
parameters corresponding to the GUT scale (Hubble scale $H_{\rm GUT}$,
coupling $g_1$ and $\alpha_{\rm GUT}$), one would obtain the same
result as above. This is because the higher cross-section at the GUT
scale $\Sigma_{\rm tot}\propto \alpha_{\rm GUT}$, compensates for the
smaller Hubble scale $H_{\rm GUT}$. Overall, we estimate the
uncertainty in the calculation of $Y^s_{3/2}$ to be  $\la 1$
order of magnitude, and the final $Y_{3/2}$ exceeds by far the bounds
imposed by BBN, similarly to moduli.

\subsubsection{Monopoles}
\label{subsubsec3.3.4}

Finally, it is important to mention that the PBB scenario also suffers
from the usual monopole problem due to GUT symmetry breaking (see also
Ref.~\cite{BM99}).  Assuming that $p$ monopoles form per horizon
volume $\sim (4\pi/3) H_C^{-3}$ at GUT symmetry breaking, one finds
that the density parameter in monopoles today is

\beq \Omega_{\rm M}h^2 \sim 10^{11}\,p\, \left (\frac{m_M}{10^{16}
\,\GeV} \right )\, \left (\frac{T_C}{10^{14}\,\GeV} \right )^3
\,\Omega_r^{-1/2}\,, \eeq
where $T_C\sim 0.5\Omega_r^{1/4}H_c^{1/2}\mpl^{1/2}$ is the critical
temperature of the phase transition, $m_M$ is the monopole mass, and
$h$ denotes the Hubble constant today in units of
100km/s/Mpc. Naively, one expects $p\sim 1/8$ by counting the number
of field orientations per horizon volume that would give rise to
monopoles. However if the radius of nucleated bubbles at coalescence
is much smaller than the horizon volume, one could actually obtain
$p>1$~\cite{VS94}.

\section{Entropy production and baryogenesis}
\label{sec4}

The previous section indicated the need for a major source of
entropy production in PBB models without an intermediate phase of
dynamics.  This is a stringent requirement, but, as we discuss below,
sufficient entropy can be produced to solve the
moduli/gravitino/monopole problems. Furthermore, as we argue in
Section 4.B, this provides a natural framework for implementing
baryogenesis in the PBB scenario.

\subsection{Sources of entropy and dilution of dangerous relics}
\label{subsec4.1}

The late decay of non-relativistic matter is a simple way to generate
entropy.  Consider in addition to the radiation background the
presence of matter with an equation of state $p_1=w\rho_1$ and $w
<1/3$. Let us denote the value of the scale factor at the time the
energy density $\rho_1$ is equal to the radiation density, $\rho_0$,
by $a_{\rm dom}$ corresponding to a Hubble scale $H_{\rm dom}$. For $a
> a_{\rm dom}$, the Universe will be dominated by $\rho_1$ until its
decay at $a_{\rm RH}$ corresponding to a Hubble scale $H_{\rm RH}$.
To show the explicit dependence on the scale factor, let us write
$\rho_0 = g_{\star \rm dom} {\tilde \rho_0} /a^4$ where $g_{\star\rm
dom}$ denotes the number of degrees of freedom at $a_{\rm dom}$ and
$\rho_1 = {\tilde \rho_1} /a^{3(w+1)}$. Then at $a_{\rm dom}$ we have,
$g_{\star \rm dom} {\tilde \rho_0} = {\tilde \rho_1} a_{\rm
dom}^{4-3(w+1)}$.  Assuming instantaneous decay, we can denote the
energy density of radiation produced in the decay by, $\rho_2 =
g_{\star \rm RH} {\tilde \rho_2} /a^4$ where $g_{\star\rm RH}$ denotes
the number of degrees of freedom at reheating and $g_{\star \rm RH}
{\tilde \rho_2} = {\tilde \rho_1} a_{\rm RH}^{4-3(w+1)}$.

If we call $s_{0}$, the entropy density contained in $\rho_0$ at
$a_{\rm RH}$, then $s_0 = {4\over 3} g_{\star \rm dom} {\tilde
\rho_0}^{3/4} /a_{\rm RH}^3$.  Similarly, the entropy in the radiation
produced by the decay is $s_1 = {4\over 3}g_{\star \rm RH} {\tilde
\rho_2}^{3/4} /a_{\rm RH}^3 ={4\over 3} g_{\star \rm RH}^{1/4} {\tilde
\rho_1}^{3/4} a_{\rm RH}^{-9(w+1)/4} = {4\over 3}g_{\star \rm RH}^{1/4}
g_{\star \rm dom}^{3/4} {\tilde \rho_0}^{3/4} a_{\rm RH}^{-9(w+1)/4} /
a_{\rm dom}^{ [3 -9(w+1)/4]}$.  If we assume that the entropy release
is large, we can write

\beq
\Delta s \equiv {s_1 - s_0 \over s_0} \simeq {s_1 \over s_0}  = 
\left({g_{\star \rm RH}
\over g_{\star{\rm dom}}}\right)^{{1 \over 4}} 
\left( { a_{\rm RH} \over a_{\rm dom}} \right)^{[3-9(w+1)/4]} 
\Omega_r^{-3/4}
\label{ds}
\eeq
We can also express the entropy change in terms of the Hubble
parameter using $(H_{\rm RH}/H_{\rm dom})^2 = ({ a_{\rm dom} /
a_{\rm RH}})^{3(w+1)}$ so that
\begin{equation}
\Delta s = \left (\frac{g_{\star \rm RH}}{g_{\star \rm dom}} \right )^{1/4}
\,\left (\frac{H_{\rm dom}}{H_{\rm RH}}\right )^{[4-3(1+w)]/[2(1+w)]}\,
\Omega_r^{-3/4}.
\label{ds2}
\end{equation}
Note that we included explicitly in $s_{\rm dom}$ a factor of
$\Omega_r^{3/4}$, which accounts for the fact that the FRW era may be
driven by relativistic fields, but not by radiation (meaning gauge
fields of the observable sector).  We also assume that the dilaton is
fixed to its present value, at the latest by the time of domination.

Depending on the equation of state, the exponent $[4-3(1+w)]/[2(1+w)]$
takes values from $1/2$ for $w=0$ (non-relativistic matter) to
$\to+\infty$ for $w\to-1$ (cosmological constant), which is what
effectively happens in standard inflation. Entropy can also be
produced in first order phase transitions, albeit to a modest level,
generally not more than $\sim1$ order of magnitude~\cite{EENO}.

In the following, we will be interested in the case of domination and
decay of oscillations of a classical scalar field $\chi$ in its
potential $V(\chi)$. In the present scenario we will assume that
initially the field $\chi$ is displaced from its low-energy minimum by
an amount $\chi_0 \sim \Ms \sim \mpl$.  This assumption is reasonable
so long as the energy scales we are considering are much larger than
the mass of the scalar field.  If we stick with canonical kinetic
terms for the moduli during the PBB phase and we appeal to no-scale
supergravity models to describe the particle content at the beginning
of the radiation era, the flat directions corresponding to the moduli
are still preserved, at least at tree-level~\cite{CGMO98}.  When
supersymmetry breaking occurs the moduli will get a mass and we
assume that the potential takes the simple form $V(\chi)\simeq
m_\chi^2\,\chi^2/2$.

The dynamics of a scalar field in its potential in the expanding 
Universe are well-known: the field is
overdamped, and remains frozen to its initial value $\chi_0$ as long
as $H\ga m_\chi$. For $H\la m_\chi$, the field oscillates with an
amplitude $\propto a^{-3/2}$. Provided $\chi_0\la \mpl$, the field
comes to dominate the energy density after having started oscillating;
if, as before, domination occurs at Hubble scale $H_{\rm dom}$, then
for $H_{\rm dom}<H\la m_\chi$, the amplitude of $\chi$ $\propto
H^{3/4}$ since the Universe is still radiation dominated; for
$H<H_{\rm dom}$, its amplitude $\propto H$. If we denote by $a_\chi$
the value of the scale factor when oscillations begin, then the
oscillations dominate at $a_{\rm dom} = (\mpl/\chi_0)^2 a_\chi$.  The
field decays when $H\sim \Gamma_\chi$, where $\Gamma_\chi$ is the
decay width of $\chi$; as before, we write $H_{\rm
RH}=\Gamma_\chi$. Assuming that $\chi$ has gravitational interactions,
$\Gamma_\chi\simeq \gamma_\chi m_\chi^3/\Mpl^2$, where $\gamma_\chi$
is a fudge factor, we find that $\chi$'s decay at $a_{\rm RH} =
\gamma_\chi^{-2/3}(\chi_0/\mpl)^{2/3} (m_\chi/\Mpl)^{-4/3}
a_\chi$. Inserting these expressions for $a_{\rm dom}$ and $a_{\rm
RH}$ into Eq.~(\ref{ds}) with $w=0$, we get

\begin{equation}
\Delta s \simeq 10^{12}\gamma_\chi^{-1/2} 
\left(\frac{m_\chi}{10^6\,{\rm GeV}}\right)^{-1} 
\left(\frac{\chi_0}{\mpl}\right)^2\Omega_r^{-3/4},
\label{dS_chi}
\end{equation}
where we have set $g_{\star\rm dom}=200$, and $g_{\star\rm RH}=10$.  

In the above, we chose to select a gravitational decay timescale for
the $\chi$ field, as it represents the most efficient source of
entropy, and $\chi$ is therefore the coherent mode of a hidden
sector scalar or modulus. In principle it is possible to obtain more
entropy production if $m_\chi\la10^6\,$GeV. However the reheating
temperature, given by
\beq
T_{\chi
\rm RH}\simeq 0.6\,{\rm GeV}\, \gamma_\chi^{1/2}\,
\left (\frac{g_{\star\rm RH}}{10} \right )^{-1/4}\, 
\left (\frac{m_\chi}{10^6\,{\rm GeV}} \right )^{3/2}\,,
\label{4.4}
\eeq
should not be lower than $\simeq 10\,$MeV for BBN to proceed
unaffected, which requires $m_\chi\ga
6\times10^4\,\gamma_\chi^{-1/3}$GeV~\cite{wein}.  Furthermore
gravitinos are re-created in $\chi$ decay to the level of: $Y_{3/2}
\simeq 10^{-13}(m_{3/2}/1\,{\rm TeV})^2(m_\chi/10^6\,{\rm
GeV})^{-3/2}$, so that one should impose $m_\chi\ga 10^5 -
10^6\,{\rm GeV}$~\cite{HIYY98}. Finally, if $R-$parity holds, one
needs to achieve $T_{\chi\rm RH}\ga 1\,$GeV for annihilations of LSPs
to take place efficiently enough to reduce its abundance to
cosmologically acceptable levels~\cite{KMY95}. Overall, it seems that
$\Delta s\sim10^{12}$ represents, within an order of magnitude, the
largest entropy production that is compatible with cosmological bounds
for a displaced oscillating modulus.

If at time $\eta_1$, $Y_m\sim 0.3g_\star^{-1}$, the final abundance of
moduli is given by:

\begin{equation}
\label{Y_mf}
Y_m \sim 2\times10^{-15}\gamma_\chi^{1/2}
\left(\frac{m_\chi}{10^6\,{\rm GeV}}\right)
\left(\frac{\chi_0}{\mpl}\right)^{-2}\Omega_r^{3/4} \, ,
\end{equation}
where we assumed, for simplicity, $g_\star\simeq200$ for the number of
degrees of freedom in the radiation bath at time $\eta_1$, and
$\Omega_r$ represents as before the fraction of energy density stored
in particles charged under gauge groups. For higher $g_\star$, the
numerical prefactor is further reduced as $(g_\star/200)^{-1}$.  We
found in the previous section that the abundance of gravitinos
produced gravitationally and in scattering processes does not exceed
the abundance of moduli, and therefore the above estimate provides an
upper limit to the final abundance of gravitinos. As for the monopole
density parameter today, it is given by: 

\beq
\Omega_M h^2\simeq
0.1p\,\gamma_\chi^{1/2}\,
\left (\frac{m_M}{10^{16}\,\GeV}\right )
\,\left (\frac{T_C}{10^{14}\,\GeV}\right )^3
\,\left (\frac{m_\chi}{10^6\,{\rm
GeV}}\right )\,\left (\frac{\chi_0}{\mpl}\right )^{-2}\,\Omega_r^{1/4}\,.
\label{mon}
\eeq
This result was already obtained 
in Ref.~\cite{BM99}, which studied the dilution of the monopole abundance
through moduli oscillations and decay.

Therefore, an initial value $\chi_0\sim \mpl$ appears sufficient to
solve the moduli/gravitino problem of the PBB scenario, and marginally
sufficient with regards to the monopole problem. If
$\chi_0\sim10^{17}\,$GeV, it is still possible to dilute the moduli
and gravitinos down to acceptable levels $Y_m\sim 10^{-13}$, but not
monopoles. It should be pointed out, however, that the number $p$ of
monopoles produced per horizon volume is uncertain, and furthermore,
that annihilations of monopoles with antimonopoles have been neglected
in the above calculations. As a matter of fact, in various patterns of
symmetry breaking, it appears that the monopoles are tied by cosmic
strings, in which case annihilation of monopoles and antimonopoles
would be highly efficient, leading to a scaling regime with one
monopole per horizon volume {\it at all times}~\cite{VS94,CHKMT88}. In
this latter case, there would be no monopole problem at all.

It should also be noted that we did not mention the possible moduli
problem associated with the coherent mode of those moduli whose mass
$\la 10\,$TeV, even though we considered entropy production due to one
such coherent mode with mass $\sim 10^6\,$GeV.  As is well-known, an
initial displacement of order $\mpl$ from the low-energy minimum of
these moduli potentials would lead to a cosmological catastrophe: a
reheating temperature $\la 1\,$keV and an enormous post-BBN entropy
production $\ga10^{16}$~\cite{CFKRR83}. Once the modulus starts
oscillating, it behaves as a condensate of zero momentum particles
with abundance $n_m/s\sim 10^7 (m/100\,{\rm
GeV})^{-1/2}(\chi_0/\mpl)^2\Omega_r^{-3/4}$, where $m$, $\chi_0$,
$\Omega_r$ and $s$ denote respectively the mass, initial vev and
radiation fraction of energy density when the modulus starts its
oscillations, and the entropy contained in radiation.  Hence, the
above source of entropy cannot reduce sufficiently the abundance of
these moduli, unless $\chi_0\la 10^{-4}\mpl$. Unfortunately there is
no well accepted reason why at high energy, i.e. after the branch
change, these moduli should lie close to their low-energy minimum, and
this problem affects all cosmological models, not only the PBB
scenario.  Nevertheless, if the string vacuum is a point of enhanced
symmetry, one would indeed expect the moduli to lie close to their
low-energy minima at high energy scales~\cite{DRT95}. In this respect,
there is an interesting difference between the PBB scenario and the
standard inflationary models. In effect, in this latter class of
models, even if the coherent mode is not displaced at the classical
level, the generation of quantum fluctuations on large wavelengths
will generate an effective zero-mode on the scale of the horizon at
the end of inflation, displaced from its low-energy minimum: the
so-called quantum version of the moduli problem
~\cite{GLV84,GRT99a}. In the PBB scenario, since the spectrum of the
fluctuations of a scalar field is very steep at very large wavelengths
(spectral slope $= 3$), its amplitude is small enough not to
regenerate, at a quantum level, the zero-mode moduli problem.

Enhanced symmetry on the ground state manifold
would not apply to the dilaton~\cite{DRT95}, and in this respect one
could wonder whether the dilaton could not play the role of the $\chi$
field above, while other moduli would be fixed to their low-energy
minimum for the above reason. Since the dilaton drives the DDI phase
with its kinetic energy, and since it is not expected to lie exactly
at the minimum of its low-energy potential at the end of DDI, this
possibility seems rather natural in the framework of the PBB
scenario. Furthermore, it should be noted that indeed, in some
realizations of gaugino condensation, the dilaton acquires a mass as
high as $\sim10^6\,$GeV~\cite{BGW97}.

Finally, another solution to the cosmological moduli problem involves
thermal inflation \cite{LS95}, or more generally a secondary short
stage of inflation at a low scale.  Indeed if this period of inflation
takes place at a scale $H\ll m_{3/2}$, the effective potential of the
modulus during inflation will correspond to its low-energy potential
(i.e in the vacuum of broken supersymmetry), and the modulus will be
attracted exponentially fast to its minimum.

\subsection{Baryogenesis}
\label{subsec4.2}

The above source of entropy comes with a bonus, namely baryogenesis
can be implemented in a natural way via the Affleck-Dine (AD)
mechanism~\cite{AD}.  As already discussed above, the string model we
are implementing has many flat directions.  Generically, in these
vacua, the scalar quarks and leptons have non-zero expectation values
and can be associated with a baryon number and CP violating operators.
Supersymmetry breaking lifts the flat directions providing a mass to
the condensate made of squarks and sleptons, the so called AD
condensate.  When the expansion rate of the Universe is of the order
of the mass of the AD field, this field starts to oscillate coherently
along the flat directions carrying the baryon number. Finally, the
subsequent decay of the AD condensate generates the baryon asymmetry.

As it will be useful for the subsequent discussion, we will briefly
outline how this mechanism works. Let us denote by $\Phi$ the AD
condensate, and $m_\Phi$ its mass. We assume that $\Phi$ is initially
displaced by an amount $\Phi_0$ from its low-energy potential, for
reasons similar to those previously discussed. The energy density
stored in the condensate is simply $\rho_\Phi = m_\Phi^2
\Phi^2$. $\Phi$'s begin to oscillate when $H \sim m_\Phi$ at $a =
a_\Phi$ and come to dominate the expansion when $a = a_{\rm dom} =
(\mpl/\Phi_0)^2a_\Phi$.  After oscillations begin the amplitude of the
oscillations decreases as $a^{-3/2}$.  The decay width of the
condensate can be written as $\Gamma_\Phi\sim \gamma_\Phi
m_\Phi^3/\Phi^2$, where $\Phi$ is the time-dependent amplitude of
$\Phi$, and $\Phi\ll\mpl$ at decay; $\gamma_\Phi\sim \alpha^2/4\pi$ is
a fudge factor. Thus the condensates decay when $a = a_{d\Phi} =
(\Phi_0^{2/3}/\mpl^{2/9} \gamma_\Phi^{2/9} m_\Phi^{4/9}) a_\Phi$. This
is true so long as the Universe is dominated by $\Phi$ oscillations at
the time of their decay, thus requiring that $\Phi_0 > m_\Phi^{1/6}
\mpl^{5/6}$.  

The baryon number stored in the condensate oscillations is given by
\beq n_{\rm B} \simeq \epsilon {\lambda_\Phi}\,{m_\Phi}^{-1}\,
\Phi_0^2\,\Phi^2\, = \epsilon {\lambda_\Phi}\,{m_\Phi}^{-1}\,\Phi_0^4
\left( {a_\Phi \over a} \right)^3\,, \eeq where $\lambda_\Phi$ is an
effective B-violating quartic coupling.  The entropy produced
subsequent to the decay of the AD condensates is roughly $s \simeq
g_{\star{\rm RH}}^{1/4}\rho_\Phi^{3/4}$, so that the produced
baryon-to-entropy ratio is

\beq {n_B \over s} = g_{\star{\rm RH}}^{-1/4}{\epsilon\,
\lambda_\Phi\, \Phi_0^{5/2} \over m_\Phi^{5/2}} \left ({a_\Phi \over
a_{d\Phi}} \right )^{3/4} = g_{\star{\rm RH}}^{-1/4}\epsilon\,
\lambda_\Phi\, {\Phi_0^2 \over m_\Phi^2} \left ({\gamma_\Phi\, \mpl
\over m_\Phi} \right )^{1/6}\,,
\label{nbs}
\eeq where $\epsilon$ is a CP-violating phase.  As entropy is produced
in the AD condensate decay, the abundance of moduli and gravitinos
produced at the end of the PBB era will be diluted.  Using
Eq.~(\ref{ds}) above, one can easily show that $\Delta s$ is given by

\beq 
\Delta s = \left (\frac{g_{\star {\rm RH}}}{g_{\star{\rm
dom}}}\right )^{1/4}\,\gamma_\Phi^{-1/6}\,\Omega_r^{-3/4}\,{\Phi_0^2 \over
m_\Phi^{1/3}\, \mpl^{5/3}}\,, 
\label{eAD}
\eeq 
and is only ${\cal O}(10^6)$ for $m_\Phi \sim 100$ GeV; the
corresponding reheating temperature is 
\beq
T_{\rm RH}\simeq 10^5\,{\rm
GeV}\,g_{\star\rm RH}^{-1/4}\,\gamma_\Phi^{1/6}\,\left (
\frac{m_\Phi}{100\,{\rm
GeV}}\right )^{5/6}\,.
\label{4.10}
\eeq 
Thus we see that unfortunately, the entropy produced in
the decay of the squark-slepton condensate cannot by itself produce
the required source of entropy.

As one can easily see from Eq. (\ref{nbs}), the Affleck-Dine scenario
of baryogenesis tends to produce too large a baryon asymmetry, with
$n_{\rm B}/s\sim{\cal O}(1)$ if the AD condensate dominates the
evolution when it decays. One possibility to reduce the baryon
asymmetry that is of interest in the present context, is the late
entropy production, as pointed out in
\cite{EENO,CGMO98,MYY94}. Moreover, it turns out that in the
cosmological scenario envisaged here, the Affleck-Dine scenario seems
to be the only model of baryogenesis capable of producing the required
baryon asymmetry. Indeed, since $\Delta s\ga10^{10}$ is required, and
since BBN indicates a baryon asymmetry $n_{\rm B}/s\sim 4 -
7\times10^{-11}$, if baryogenesis takes place before entropy
production, one needs to achieve $n_{\rm B}/s\sim {\cal O}(1)$
initially, and only the Affleck-Dine mechanism seems capable of such a
feat.

Let us now consider the combined effect of an AD condensate and the
late decay of a moduli field.  Since $m_\Phi\sim{\cal O}(100\,{\rm
GeV})$, the $\chi$ field above will start oscillating before $\Phi$.
In the PBB scenario with no intermediate phase, the value of the
Hubble parameter at the end of the PBB phase is $H_1 \sim M_s$ at $a=
a_1$ and it is much larger than the value of $H$ when $\chi$ would
start to oscillate, which happens at $H \sim m_\chi \sim 10^6\,{\rm
GeV}$.  Hence, before $\chi$ starts to oscillate the Universe is in a
radiation dominated era. 

As before, one can determine the epoch of $\chi$ domination (at scale
factor $a_{\rm dom}$) by setting $\rho_r = \rho_\chi$, using
$(a_\chi/a_1)^2 = H_1/m_\chi$ ($a_\chi$ is the value of the scale
factor when $\chi$'s begin to oscillate), giving:
${a_1/a_{\rm dom}} = {m_\chi^{1/2}\chi_0^2/H_1^{1/2}\mpl^2}$.
An analogous equation can be written for $\Phi$ and we see that
provided $m_\Phi < 10^6\,{\rm GeV} (m_\chi/10^6\,{\rm
GeV})(\chi_0/\Phi_0)^4$, $\chi$ will dominate the energy density
before $\Phi$ ($a_{\rm dom}^\Phi > a_{\rm dom}^\chi$) and this
condition is satisfied for most values of the parameters we are
interested in.  Moreover $\Phi$ decays before $\chi$.  Therefore,
since $\chi$ dominates the evolution before $\Phi$, and since $\chi$
decays after $\Phi$, the total entropy produced remains the same as in
Eq.~(\ref{dS_chi}) above and is sufficient for solving the moduli and
gravitino problems of the PBB scenario.

We will assume that the $\chi$ field dominates the energy density
before $\Phi$ begins to oscillate at $a_\Phi$; we will relax this
assumption further below.  This condition is true as long as $m_\Phi <
m_\chi (\chi_0/\mpl)^4$, which for $m_\Phi \sim 10^2\,{\rm GeV}$ and
$m_\chi \sim 10^6\,{\rm GeV}$, gives $\chi_0 > 10^{-1}\,\mpl$, which
is quite reasonable in our context. Thus we can relate $a_\chi$ and
$a_\Phi$ through $a_\chi/a_\Phi =
(m_\Phi/m_\chi)^{2/3}(\chi_0/\mpl)^{-2/3}$.  In this case, we can
rewrite the baryon number stored in the condensate oscillations in
terms of $a_\chi$ as
\beq
n_{\rm B} \simeq  \epsilon
{\lambda_\Phi}\,{m_\Phi}^{-1}\,\Phi_0^4 \left( {a_\Phi \over a} \right)^3
= \epsilon {\lambda_\Phi}\, \Phi_0^4\, {m_\chi^2\, \chi_0^2 \over
{m_\Phi^3} \mpl^2}\,  \left( {a_\chi \over a} \right)^3\,,
\eeq 
As before, we will consider the gravitational decay of $\chi$ to
proceed with a rate $\Gamma_\chi = \gamma_\chi m_\chi^3
/\Mpl^2$. $\chi$'s decay at $a_{\rm RH}$ when $H = \Gamma_\chi$ or
when $(a_\chi/a_{\rm RH})^3 = \gamma_\chi^2
m_\chi^4\chi_0^{-2}\mpl^{-2}/(8\pi)^2$.  Subsequent to decay, the
Universe reheats to $g_{\star \rm RH} T_{\rm RH}^4=\rho_\chi(a_{\rm
RH})=m_\chi^2\chi_0^2(a_\chi/a_{\rm RH})^3$ and the entropy density is
just $(4/3) \rho_\chi /T_{\rm RH}$. Thus the baryon to entropy ratio
is easily determined to be \cite{CGMO98} 

\beq
\frac{n_{\rm B}}{s}\simeq 0.1
\epsilon \lambda_\Phi\,\gamma_\chi^{1/2}\,g_{\star
\rm RH}^{-1/4}\, { \Phi_0^4 \, m_\chi^{3/2}\, \over m_\Phi^3\,
\mpl^{5/2}}\,,
\label{nb_s}
\eeq
For an effective quartic coupling $\lambda_\Phi\sim 
m_\Phi^2/(\Phi_0^2+M_X^2)$, corresponding to superheavy gaugino
exchange of mass $M_X$, 
one finds: 

\begin{equation}
\frac{n_{\rm B}}{s}\simeq 3.\times 10^{-4}\epsilon\,\gamma_\chi^{1/2}\,
\left(\frac{m_\Phi}{100\,\GeV}\right)^{-1}\,
\left(\frac{\Phi_0}{\mpl}\right)^2\,
\left(\frac{m_\chi}{10^6\,\GeV}\right)^{3/2}\,
\frac{\Phi_0^2}{\Phi_0^2+M_X^2}\,,
\label{nb2s}
\end{equation}
and we assumed $g_{\star{\rm RH}}=10$ for simplicity. 
 
If the $\Phi$ field starts to oscillate before the $\chi$ field 
dominates, i.e. if $\chi_0 < 10^{-1}\,\mpl$, 
then the r.h.s. of Eq.~(\ref{nb2s}) should be multiplied by
$(m_\Phi/m_\chi)^{1/2}(\chi_0/\mpl)^{-2}$. The baryon to entropy 
ratio in this case is:

\begin{equation}
\frac{n_B}{s}\simeq2\times10^{-5} \epsilon
\,\gamma_\chi^{1/2}\,
\left(\frac{g_{\star \rm RH}}{10}\right)^{-1/4}\,
\left(\frac{m_\chi}{10^6\,{\rm GeV}}\right)\,
\left(\frac{m_\Phi}{100\,{\rm GeV}}\right)^{-1/2}\,
\left(\frac{\Phi_0}{\mpl}\right)^2\,
\left(\frac{\chi_0}{\mpl}\right)^{-2}\,
\frac{\Phi_0^2}{\Phi_0^2+M_X^2}.
\label{nb1s}
\end{equation}
Both values of $n_{\rm B}/s$ obtained, Eqs.~(\ref{nb2s}) and
(\ref{nb1s}), are too large, but as discussed in Ref.~\cite{CGMO98},
it is likely to be reduced through various mechanisms.  For instance,
if baryon number violation is Planck suppressed, then
$\lambda_\Phi\sim m_\Phi^2/\mpl^2$, which would reduce the above
asymmetry if $\Phi_0\la \mpl$. Furthermore, under certain conditions,
non-renormalizable interactions may reduce the initial vev of the
condensate down to possibly $M_X\sim 10^{16}\,$GeV, and the baryon
asymmetry would be reduced by $\simeq2 - 3$ orders of
magnitude~\cite{CGMO98}. Finally sphaleron processing of the baryon
asymmetry in the electroweak phase transition can also lead to
reduction of $n_{\rm B}/s$, by as much as $\simeq6$ orders of
magnitude.  Therefore, it seems reasonable to conclude that a baryon
asymmetry of the right order of magnitude can be produced in the
Affleck-Dine mechanism in this scenario.

\vskip 0.5truecm Let us now discuss the implications of the presence
of an intermediate phase between DDI and FRW on the above
conclusions. As we shall see, one of the main features is that the
``democracy rule'' of gravitational particle production no longer
necessarily applies in the presence of an intermediate phase, and the
distribution of energy density among the various components may be
drastically altered. In some cases, this will imply that gravitinos
and moduli can be more easily diluted in entropy production.

\section{Dual-dilaton intermediate phase}
\label{sec5.1}

The dual-dilaton era is characterized by two wavenumbers $k_1$ and
$k_s$ that correspond to the horizon size at the branch change
between the ``dual-dilaton'' intermediate phase and FRW, and between
DDI and the ``dual-dilaton'' era, respectively.  During the
intermediate phase (IP), modes re-enter the horizon [since wavelengths
$\propto a(\eta)$ do not increase as fast as the horizon size
$H^{-1}=a^2/a'$], and therefore $k_1 < k_s$.  As a consequence, one
still expects to produce roughly one particle per mode at the highest
wavenumber $k_s$, since it exited and re-entered simultaneously at
time $\eta_s$. We will also assume $H_s\sim M_s$, since $H_s$
correspond to the maximal value of the Hubble scale. At times $\eta >
\eta_1$, one finds an expression similar to Eq.~(\ref{A11}) for the
energy density of species $j$:

\bea
\label{B2}
\frac{d \rho_j}{d \log k}(\eta) &\simeq& \frac{{\cal N}_j}{2 \pi^2}\,
H_s^4\,\left ( \frac{a_s}{a} \right )^{4}\, \left ( \frac{k}{k_1}
\right )^{n_j^{\rm \tiny DDI}}\, \left ( \frac{k_1}{k_s} \right
)^{n_j^{\rm \tiny IP}} \quad k_{\rm H} \ll k \ll k_1 \\
\label{B3}
& \simeq & \frac{{\cal N}_j}{2 \pi^2}\, H_s^4\,
\left(\frac{a_s}{a}\right)^{4}\, \left ( \frac{k}{k_s} \right
)^{n_j^{\rm \tiny IP}} \quad k_1 \ll k \ll k_s \,, \eea where
$n_j^{\rm \tiny IP}$ is the spectral slope for fluctuations that
exited the horizon during DDI and reentered during the ``dual-dilaton``
intermediate era. We again
imposed a low wavenumber cut-off $k_{\rm H}$ corresponding to the
horizon size, but it will not play a role for $\eta > \eta_1$, since $
n_j^{\rm \tiny DDI}>0$ as before for all fields. The integrated energy
density can be written as:

\beq \rho_j(\eta) \simeq \frac{{\cal N}_j}{2 \pi^2}\, H_s^4\,\left (
\frac{a_s}{a} \right )^{4}\, \left [ 1 +
\left(\frac{k_s}{k_1}\right)^{-n_j^{\rm \tiny IP}}\right ]
\quad\quad\eta > \eta_1\,.  
\label{B4}
\eeq 
\subsection{Particle content and reheating} 

{}From Eq.~(\ref{B4}) we obtain that at times $\eta > \eta_1$, 
the energy density is dominated by
the field with the most negative spectral slope $n_j^{\rm \tiny IP}$,
and the democracy rule does not apply, unless all $n_j^{\rm \tiny
IP}>0$. The spectral slopes of the various fields depend in a
non-trivial way on the dynamics of the internal dimensions during DDI
and during the ``dual-dilaton'' intermediate era (see Tab.~\ref{tab1}). 
Since details can be found in Ref.~\cite{BMUV98}, we
will simply restate the relevant results, but extend the analysis 
to the case with dynamical internal dimensions during the IP, i.e. 
$\theta \neq 1/\sqrt{3}$.

\begin{figure}[c]
\epsfysize=3truein
$$\epsfbox{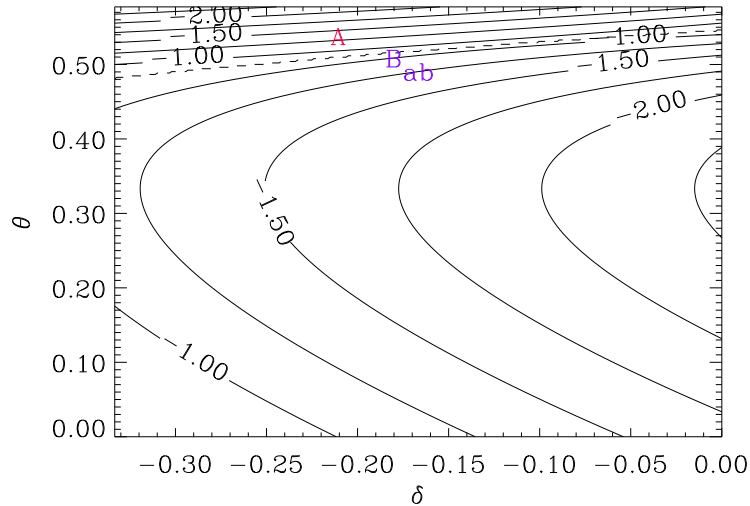}$$
\caption{The contours delimit the values of the smallest spectral
slope ${\rm min}_j(n_j^{\rm IP})$ as a function of the parameters
$\delta$ and $\theta$, defined in Eqs.~(\protect\ref{A8}),
(\protect\ref{A10}). The dashed line separates the regions 
in which either the model-independent axion $A$ or 
the axion $B_{ab}$ dominates.}
\label{Fig1}
\end{figure}

In Fig.~\ref{Fig1}, we show a contour plot of the smallest spectral
slope ${\rm min}_j(n_j^{\rm\tiny IP})$ in the parameter space
$\delta-\theta$; we recall that $\delta$ and $\theta$ characterize the
evolution of external dimensions during DDI and the IP respectively
[see Eqs.~(\ref{A8}), (\ref{A10})]. The dashed line separates the
areas which refers to the model-independent axion $A$, and to the
axion $B_{a b}$ associated to the internal components of
$B_{\mu\nu}$. The spectral slope of these latter depend sensitively on
the evolution of the 6 internal dimensions during DDI and IP, and in
Fig.~\ref{Fig1}, we chose to show the case in which $2$ internal
dimensions compactify isotropically. In the following, we will examine
the various cases in which $a$ dimensions compactify isotropically,
and the other $6-a$ are stabilized during DDI and IP. We find that the
following two possiblities arise:

{\bf (1)} if $|\delta| \sim 1/3$ and $\theta$ is close to its maximal
value, which corresponds to saturation of the Kasner constraint
Eq.~(\ref{A9}), i.e. stabilized internal dimensions, then the smallest
slope is negative, and is carried by $A$. In this case, the
model-independent axion carries all the energy density at
$\eta>\eta_1$. However, when one assumes that compactification is
anisotropic, i.e. $a $ dimensions compactify isotropically
and $6-a$ are stabilized during both DDI and IP one obtains a
slightly different picture. In some cases, as the one shown 
in Fig.~\ref{Fig1}, the minimal slope is
everywhere negative, and the bulk of the energy density is carried by
either $A$ or $B_{ab}$ axions, depending on the value of $\delta$ and
$\theta$.

{\bf (2)} if the smallest spectral slope is positive, meaning
$n_j^{\rm\tiny IP}>0$ for all $j$, then all fields roughly share the
same energy density at time $\eta>\eta_s$. This case is therefore very
similar to that envisaged in Section \ref{sec3}, i.e. for a sudden branch
change. Notably, one finds $Y_m\sim 0.3/g_\star$ as previously, and
entropy production as before may be considered to eliminate this
problem. We will thus ignore this case in the following, and rather
concentrate on case {\bf (1)}, assuming that one of the axions
dominate the energy density at time $\eta_1$.

In the following, we denote generically as ${\cal A}$ the axion field
that carries the energy density at times $\eta>\eta_1$, and when
necessary we will specify whether ${\cal A}$ is $A$ or $B_{ab}$. The
fraction of energy density contained in species $j$ is:

\beq
\Omega_j \simeq {\cal N}_j\left|\frac{\eta_1}{\eta_s}\right|^{n_{\cal
A}^{\rm \tiny 
\tiny IP}}
\left[ 1 + \left|\frac{\eta_1}{\eta_s}
\right|^{-n_j^{\rm \tiny \tiny IP}}
\right],\quad\quad \eta>\eta_1
\label{rB5}
\eeq where we used $k_1/k_s = \eta_s/\eta_1$.  One can derive a
relation between the duration of the ``dual-dilaton'' phase
$|\eta_1/\eta_s|$ and the coupling constant $g_1$ at time $\eta_1$
from the criticality condition $\rho_{\cal A}\simeq\rho_c$, which gives:

\beq
\left|\frac{\eta_1}{\eta_s}\right|^{\epsilon_{\cal A}} \simeq g_1,
\label{B5}
\eeq where the exponent $\epsilon_{\cal A}= (2\theta-1)/(1-\theta) + n_{\cal
A}^{\rm \tiny \tiny IP}/2 $ and is negative in the region of the
parameter space where $ n_{\cal A}^{\rm \tiny \tiny IP}<0$.  To derive
Eq.~(\ref{B5}) we used $a(\eta)\propto |\eta|^{\theta/1-\theta}$,
$H\propto |\eta|^{-1/1-\theta}$ and
$g\propto|\eta|^{(3\theta-1)/2(1-\theta)}$, Eq.~(\ref{B4}), and
$\rho_c=(3/8\pi)g_1^{-2}H_1^2M_s^2$ at $\eta=\eta_1$. If ${\cal A}$ is
the model-independent axion $A$, $n_{\cal A}^{\rm \tiny \tiny IP}
=4/(1-\delta) - 4\theta/(1-\theta)$, while if ${\cal A}$ is a $B_{ab}$
axion, $n_{\cal A}^{\rm \tiny \tiny IP} =4 - 4|\beta_a/(1-\delta) -
\xi_a/(1-\theta)|$, where $\beta_a$ and $\xi_a$ parametrize the
evolution of the internal scale factor during DDI and the
``dual-dilaton'' phase, respectively; $\beta_a$ is tied to $\delta$ by
the Kasner constraint Eq.~(\ref{A9}) and similarly for $\xi_a$ as a
function of $\theta$ (see Ref.~\cite{BMUV98}).  Therefore
Eq.~(\ref{rB5}) can be re-written as: $\Omega_j \simeq {\cal N}_j
g_1^{n_{\cal A}^{\rm \tiny \tiny IP}/\epsilon_{\cal A}} [1 + g_1^{-n_j^{\rm
\tiny \tiny IP} /\epsilon_{\cal A}}]$. It is difficult to give quantitative
estimates for the total fraction of energy density carried by
radiation, since the various components of radiation have different
spectral slopes, and some of them can be negative (see Tab.~\ref{tab1}). 
In the following,
we thus assume that radiation can be considered, on average, as one
species with number of degrees of freedom ${\cal N}_r\sim g_\star/2$
(as before), with a positive spectral slope, which implies 

\beq
\Omega_r \simeq {\cal N}_r\, g_1^{n_{\cal A}^{\rm \tiny 
IP}/\epsilon_{\cal A}}\,.
\label{l}
\eeq Note that this number can actually be of order 1 if ${\cal N}_r$
is sufficiently large as compared to $g_1^{n_{\cal A}^{\rm \tiny \tiny
IP} /\epsilon_{\cal A}}$; however, this case would be similar to a sudden
branch change, since $g_1^{1/\epsilon_{\cal A}}\simeq|\eta_1/\eta_s|$ would be
of order unity, and the results of previous sections apply. We thus
assume $\Omega_r\ll1$ in what follows. To be definite, let us take
$g_1\simeq10^{-1}$, $\theta=1/\sqrt{3}$ (corresponding to stabilized
internal dimensions during IP) we find, varying $\delta$ in the range
$(0,-1/3)$: \beq \frac{H_1}{M_s} \sim 10^{-7}- 10^{-3}\,, \quad \quad
\Omega_r\sim {\cal N}_r\,(10^{-4}-10^{-3})\,,
\label{l1}
\eeq
while posing $g_1\simeq10^{-3}$ we get:
\beq 
\frac{H_1}{M_s} \sim 10^{-20}-10^{-8}\,, \quad \quad 
\Omega_r\sim {\cal N}_r\,(10^{-12}-10^{-9})\,, 
\label{l2}
\eeq

In contrast to the scenario with no intermediate phase, when a
dual-dilaton intermediate era is present, reheating cannot be provided
by gauge non-singlets, because at $\eta > \eta_1$ they generically
carry a small amount of energy ($\Omega_j \ll 1$), as discussed above.
Let us then investigate the possibility of reheating via the axion
fields present in our PBB model.  Reheating may proceed if the axion
can recreate radiation by scattering or conversions with photons, that
is through the processes ${\cal A} + {\cal A} \rightarrow \gamma +
\gamma$ and ${\cal A} + \gamma \rightarrow {\cal A} + \gamma$.  The
interaction rate of the latter channel is strongly suppressed relative
to the the rate of the former, since the radiation number density is
small. The interaction term between ${\cal A}$ and the gauge fields is
of the form $({\cal A}/M'){\cal F}\widetilde{\cal F}$. The mass scale
$M'\simeq\Mpl$ for the model-independent axion $A$, but $M'$ can be
different for the $B_{ab}$ axions, as it then depends on the
compactification~\cite{CK85}. Indeed, as shown in the action
Eq.~(\ref{A6}), the coupling of $B_{ab}$ to $W\widetilde{W}$ or $V
\widetilde{V}$, depends on the expectation values of the internal
moduli.

 The cross-section for ${\cal A} - {\cal A}$ scattering thus is of the
form $\sigma \sim E_{\cal A}^2/M'^4$, where the typical axion energy
$E_{\cal A}\sim H_1 (a_1/a)$, with $H_1$ possibly much smaller than
$M_s$ for a dual-dilaton intermediate phase. Finally, the axion energy
density is given by: $\rho_{\cal A}\simeq (3/8\pi) g_1^{-2}H_1^2M_s^2
(a_1/a)^4$, so that the interaction rate for scattering gives:

\beq \frac{\Gamma_{{\cal A}{\cal A}}}{H}\sim \frac{3}{8\pi}\, g_1^{2}\, \left(
\frac{H_1}{M_s}\right)^2\,\left({M'\over\Mpl}\right)^{-4}\,
 \left(\frac{a_1}{a}\right)^{3}\,.  \eeq

Therefore scattering by the model-independent axion cannot provide
reheating, as one normally expects $H_1\ll M_s$. However, if
$(M'/\Mpl) \la g_1^{1/2} (H_1/M_s)^{1/2}$, then $B_{ab}-B_{ab}$ axion
scattering will produce radiation and reheat the Universe.

If axion scattering is ineffective, reheating may still occur through
axion decay, provided the axion mass is large enough to avoid problems
associated with too low a reheating temperature. A typical axion
lifetime is $\tau_{\cal A}\sim M'^2/m_{\cal A}^ 3$, where $m_{\cal
A} \sim \Lambda^3/\Mpl M'$ is the axion mass (neglecting fudge
factors)~\cite{CK85}, and $\Lambda$ denotes the highest scale at which
gauge interactions to which ${\cal A}$ couples become strong. In particular,
for $\Lambda \sim10^{14}\,$GeV, corresponding to phenomenologically
favoured scales of gaugino condensation in a hidden sector, $m_{\cal
A}\sim 10^6 (M'/0.01\Mpl)^{-1}\,$GeV and $\tau_{\cal A}\sim 3\times
10^{-8}\,{\rm s}\,(M'/0.01\Mpl)^5$, and the reheating temperature
is $T_{\rm RH}\sim 10\,{\rm GeV}\,g_{\star{\rm
RH}}^{-1/4}(M'/0.01\Mpl)^{-5/2}$. 
In the following subsections, we will discuss the 
implications of this on the moduli and gravitino
problems.
 
\subsection{Dangerous relics}

The estimate of the abundance $Y_m$ of moduli produced gravitationally
can be obtained using the same methods as for the no-intermediate
phase case. One actually finds the same result
$Y_m\sim0.3/g_\star$. This is due to the fact that the spectral slope
of the energy distribution of moduli and radiation is positive for all
$k$ momenta (see Tab.~\ref{tab1}), and therefore the number density of
moduli $n_m\sim (2\pi)^{-3/2}H_s^{3/2}(a/a_s)^{-3}$, where we used
Eq.~(\ref{B2}), and $n_m\sim\rho_m/E_m$, with $E_m\sim
H_s(a/a_s)^{-1}$ the typical moduli energy. One can also express the
entropy density $s\sim g_\star^{1/4}\rho_r^{3/4}$ in a similar way,
and obtain the above result for $Y_m$.

Just as in the no-intermediate phase case, one does not expect
gravitinos to be produced gravitationally in the dual-dilaton phase if
they are effectively massless. However, even if radiation has
not thermalized, they can be produced by scattering during the
dual-dilaton phase. Since the string coupling (hence the production
cross-section) evolves with time during dual-dilaton phase, 
it is more convenient to
write the Boltzmann equation in terms of conformal time, which when
disregarding annihilations channels gives:

\beq \frac{d Y_{3/2}^s}{d \eta} = \Sigma_{\rm tot}\,s\,a\,Y_a^2\,,
\quad \quad \eta_s < \eta ,
\label{grav}
\eeq where we recall that the scale factor $a\propto
|\eta|^{\theta/(1-\theta)}$ during the IP, $s\propto a^{-3}$, and
$\Sigma_{\rm tot}\sim 500g^4/M_s^2$, and as before $Y_a\sim
0.3/g_\star$ denotes the ratio of number density to entropy density of
radiation quanta per helicity state. Using $g^2\propto
|\eta|^{(3\theta -1)/(1-\theta)}$ during the IP, one easily obtains as
before $Y_{3/2}^s=Y_a \Gamma/H$, with $\Gamma\sim \Sigma_{\rm tot}Y_a
s$ the gravitino production rate, and the right hand side should be
evaluated at time $\eta_1$ if $\theta>1/3$, and at time $\eta_s$ if
$\theta<1/3$.

Thus, using $(\Sigma_{\rm tot} s/H)_{\eta_1}\sim 70 g_1^{5/2}\,
g_\star^{1/4}\, \Omega_r^{3/4}(H_1/M_s)^{1/2}$, and Eq.~(\ref{l}), one
finally obtains:

\beq Y_{3/2} \sim 6g_\star^{-1}\,g_1^\kappa \,\left (\frac{H_s}{M_s}
\right )^{1/2}\,,
\label{Y_32_IP}
\eeq

with: 

\bea 
\label{w1}
&& \theta > 1/3: \quad \kappa = {5\over2} - {1\over2\epsilon_{\cal
A}(1-\theta)} + {3n^{\rm\tiny IP}_{\cal A}\over4\epsilon_{\cal
A}}\,,\\ && \theta < 1/3: \quad \kappa = {5\over2} -
{1\over2\epsilon_{\cal A}(1-\theta)} + {3n^{\rm\tiny IP}_{\cal
A}\over4\epsilon_{\cal A}} + {1-3\theta\over\epsilon_{\cal
A}(1-\theta)}\,.  \eea

Let us first observe that, if $\theta < 1/3$ the coupling $g$
decreases in time during the IP (see Eq.~(\ref{A9})).  In the scenario
under investigation, i.e. with a dual-dilaton intermediate phase, it
is assumed that quantum string-loop effects are never operative ($ g <
1$) and that because of high-curvature corrections the
Hubble-parameter makes a bounce around its maximal value at the string
scale.  If we impose that the dilaton field reaches the present value
at the end of the PBB era, unless we assume either an extremely low
decreasing of the coupling during the IP or a very short intermediate
phase (which would give for $Y_{3/2}^s$ roughly the same value as in
the scenario with no-intermediate phase), we are forced to limit to
the region of parameter space where $\theta > 1/3$.  Having restricted
ourselves to the case described by Eq.~(\ref{w1}), we find that the
exponent $\kappa$ takes values between $\sim2$ and $\sim 9$ depending
on the evolution of internal dimensions, and therefore, $Y_{3/2}\sim
10^{-4}\to10^{-10}$ for $g_1\sim 0.1$, $g_\star \sim 100$.  For
anisotropic compactification, in some region of the parameter space,
one finds even higher values of $\kappa$, for which there would be no
gravitino problem at all. One thus finds that in the presence of a
dual-dilaton phase, the gravitino is generically less efficiently
produced (possibly much less) than in the no-intermediate phase
case. This can be understood in the following way. If $\theta > 1/3$,
the string coupling grows during IP, so that the cross-section
$\Sigma_{\rm tot}\propto g^4$ is very small at the beginning of the
dual-dilaton phase, and gravitino production takes place at Hubble
scale $H_1$. However, unlike the no-intermediate phase case, here one
generically has $H_1\ll H_s\sim M_s$, and therefore gravitino
production is inefficient.  Finally, the small number density of
radiation quanta (recall $\Omega_r\ll1$) also hampers gravitino
production in this case.

\subsection{Sources of entropy and baryogenesis} 

Due to the overproduction of moduli (and possibly gravitinos), entropy
production is still necessary in the present scenario. Entropy
production is also necessary if the axion cannot reheat the Universe, e.g.
if its coupling to the gauge fields is too weak, and its mass too small.
However, there are several possible sources of entropy production, as we
now discuss.

If the axion field dominates the energy density at the beginning of
the radiation era and it reheats the Universe through scattering or
decay, reheating is accompanied by entropy production to the level of
$\Delta s\sim (\rho_{\cal A}/\rho_r)^{3/4}$, where $\rho_{\cal A}$ and
$\rho_r$ denote the energy density contained in the axion and in
radiation at reheating. If the axion is still relativistic at that
time, i.e. if it reheats the Universe through scattering at high
energy scale, then $\Delta s\sim \Omega_r^{-3/4}$, where $\Omega_r$ is
defined as before at time $\eta_1$. The moduli abundance would be
reduced to $Y_m \sim 0.3g_\star^{-1}\,\Omega_r^{3/4}$ and to obtain
$Y_m \la 10^{-13}$, one needs $\Omega_r \la 10^{-13}$. One can check
that such a low value of $\Omega_r$ cannot be obtained for realistic
values with $g_1 \,\gaq\,10^{-3}$ in the parameter space defined by
$\delta$ and $\theta$, using the results so far obtained for the axion
$B_{a b}$ (recall that the axion $A$ cannot reheat by scattering).
For example, in the case of the axion $B_{ab}$ with only 2 internal
dimensions compactified and the other 4 stabilized, the smallest value
of $\Omega_r$ for $g_1 \,\gaq\,10^{-3}$ is actually $\Omega_r \sim
10^{-6}$, with ${\cal N}_r \sim 100$, which would imply $Y_m \sim
10^{-7}$ at the end of reheating and entropy production is still
necessary. Moreover, note that $\Omega_r \sim 10^{-6}$ also
corresponds to $H_1 \sim 10^{-13} M_s$, hence $M^\prime/\Mpl \sim
10^{-8}$, which is a rather strong requirement on
$M^\prime$. Nevertheless, one should recall at this stage that the
domination and decay of an Affleck-Dine flat direction would lead to
further entropy production $\Delta s'\sim10^6$ [see Eq.~(\ref{eAD})].
Thus one could actually dilute the moduli and gravitinos down to
acceptable levels with an Affleck-Dine condensate only (no $\chi$
modulus). This would however produce too large a baryon asymmetry,
$n_{\rm B}/s\sim{\cal O}(1)$, unless the $CP$ violation parameter is
very small and/or electroweak baryon number erasure is very
efficient~\cite{CGMO98}.

Let us now consider the alternate case, in which the axion acquires a
large mass, and later decays to radiation. Entropy production then
results from the decay of the axion oscillations around the minimum of
its potential, whose generic form~\cite{CK85} is
$V({\cal A})\sim(\Lambda^6/\Mpl^2)\,\left[1-{\rm
cos}\left({\cal A}/F\right)\right]$, with $F\sim M'/\pi^2$ the axion decay
constant.  The axion vev ${\cal A}_0$ is in fact generically displaced by
$\sim M'/\pi$ from its true minimum, and its coherent oscillations and
decay will produce entropy.  Using the results of Section \ref{sec3},
one can easily obtain:

\beq \Delta s\simeq 10^{7}\, \left(\frac{{\cal A}_0}{M'}\right)^2\,
\left(\frac{M'}{0.01\Mpl}\right)^4\,
\left(\frac{\Lambda}{10^{14}\,{\rm GeV}}\right)^{-3}\,
\Omega_r^{-3/4}\,.  \eeq 

The amount of entropy produced is only marginally sufficient, but it
depends in a sensitive way on $\Lambda$ and $\Omega_r$. For smaller
values of the gaugino condensation scale, say $\Lambda\sim\,{\rm
few}\times10^{13}\,$GeV, and, say $\Omega_r\la10^{-2}$, reasonable
values of $\Delta s\ga 10^{10}$ may be achieved. Note that
baryogenesis can be implemented in the very same way as in Section III
in this context.

One should recognize that the above estimates remain somehow
qualitative, since they make particular assumptions on the axion
couplings, and therefore on the compactification process, whereas we
assumed a simple toroidal compactification. Nevertheless, our aim here
is to show that there exists various possibilities to generate entropy
to the level required. 

Even if axions cannot reheat the Universe through scattering or decay,
one may still consider the mechanism discussed in Section III, where a
modulus (the dilaton?) of mass $\sim10^6\,$GeV reheats through the
decay of its coherent oscillations. The discussion is similar to that
of Section III, up to the fact that there may be no radiation
dominated era preceeding the $\chi$ dominated era, if $H_1\la H_{\rm
dom}$ ($H_{\rm dom}$ denotes as before the Hubble scale at which
$\chi$ comes to dominate the energy density), i.e. if the dual-dilaton
era ends as $\chi$ dominates.  However, as we now argue, this does not
happen, and one always has $H_1\gg H_{\rm dom}$.  Following the
discussion of Section III to calculate $H_{\rm dom}$, using $a\propto
H^{-\theta}$ in the IP era, one finds: $H_{\rm dom}\sim m_\chi
[(8\pi/3) (g_1\chi_0/M_s)]^{2/(2-3\theta)}$, where $g_1$ is the value
of the string coupling at Hubble scale $H_{\rm dom}$, which marks the
end of IP. In the absence of the $\chi$ modulus, the transition to the
radiation dominated era would take place at $H_1 \sim H_s
g_1^{-1/\epsilon_{\cal A}(1-\theta)}$ [see Eq.~(\ref{B5})]. We
consider that in either case, the value of the string coupling at the
end of the IP, i.e. at $H_{\rm dom}$ or at $H_1$, should be close to
its present value. One then must determine whether or not $H_{\rm
dom}\la H_1$, and it can be checked that for nearly all values of the
$\delta - \theta$ parameters, we have indeed $H_{\rm dom}\ll H_1$. The
dual-dilaton phase thus ends before and radiation domination occurs
before $\chi$ dominates. All estimates made in Section III can thus be
directly applied to the present case. It is interesting to note that
here, one generically has $\Omega_r\ll1$ (the ``radiation dominated''
era is driven by ${\cal A}$), and therefore entropy production is more
efficient by a factor $\Omega_r^{-3/4}\sim 10 - 10^6$, for ${\cal N}_r
\sim 100$ and $g_1 \sim 0.001 - 0.1$.  The amount of moduli present at
$\chi$ decay is thus further reduced by this factor. The monopole
abundance today is also reduced by a factor $\Omega_r^{-1/4}\sim
2-100$ [see Eq.~(\ref{mon})]. Baryogenesis can be implemented as
before via the Affleck-Dine mechanism, and the baryon asymmetry is
given by Eq.~(\ref{nb1s}).

In fact, if $\Omega_r\la 10^{-6}$, entropy production to the level of
$\Delta s\ga10^6$ would be sufficient to dilute the moduli and
gravitinos to acceptable levels, and such entropy could be provided by
the Affleck-Dine condensate, in the absence of any $\chi$ modulus. It
should be noted, however, that such low values of $\Omega_r$ only
arise when $g_1\la 10^{-2}$.

As a conclusion, when either $A$ or $B_{a b}$ axions dominate the
energy density at the end of a dual-dilaton intermediate phase, there
are various natural sources of entropy: axion scattering, axion decay,
or domination and the decay of a modulus or an Affleck-Dine condensate.
Since one generically has $\Omega_r\ll1$, entropy production is more
efficient than in the absence of an intermediate phase, and both
moduli and monopoles can be diluted down to low levels. In some cases,
the Affleck-Dine condensate provides enough entropy by itself to solve
the moduli/gravitino problem, although one then has to cope with a
very large baryon asymmetry from the decay of the condensate.

\section{``String'' intermediate phase and black holes}
\label{sec5.2}

During the string era, modes exit and do not re-enter the
horizon. Therefore $k_s < k_1$, and one expects to produce one
particle per mode at $k_1$ since it exits the horizon and re-enters at
the same absolute value of conformal time.

However, the situation here is more delicate than in the case of the
dual-dilaton phase. Indeed, a mode that exits the horizon at conformal
time $\eta_{\rm ex}$, with $\eta_{\rm ex} <0$ in the present scenario,
will re-enter at time $\eta_{\rm re}\simeq |\eta_{\rm ex}|$ (if
$\eta_{\rm ex}\ll \eta_1$, i.e. if the mode exits well before the end
of the ``string'' phase). This means that at time $\eta_1$, which is
supposed to mark the start of the FRW regime, only those modes with
wavenumber $\sim k_1$, i.e. the highest frequencies, have
re-entered. The modes that exited the horizon at the beginning of the
``string'' phase (wavenumber $\sim k_s$) will re-enter later, possibly
much later, at conformal time $|\eta_s|\gg |\eta_1|$. One can rewrite
the energy density Eq.~(\ref{A11}) in this scenario, at time $\eta$
with $\eta_1<\eta<|\eta_s|$, which gives:

\beq 
\label{C1a}
\rho_j(\eta) \simeq \frac{{\cal N}_j}{2\pi^2}H_1^4\,\left
(\frac{a_1}{a} \right )^{4} \, \left | 1 - \left|
\frac{\eta}{\eta_1}\right|^{-n_j^{\rm \tiny IP}}\right | \quad\quad
\eta_1<\eta<|\eta_s|\,, \eeq and as before, we imposed a low
wavenumber cut-off at the horizon size $\sim 1/\eta$. In the case of
the ``string'' phase, inspection of Table~\ref{tab1} reveals that the
model-independent axion has the most negative slope $=-2\zeta$
($\zeta>0$), and its energy density $\rho_A\sim ({\cal N}_A/2\pi^2)
H_1^4 a_1^4 |\eta/\eta_1|^{2\zeta}/a^4$. At time $\eta_1$, only modes
with wavenumber $k_1$ have re-entered, so within the horizon, all
fields share roughly the same energy density. However, as time goes
beyond $\eta_1$, the axion will quickly come to dominate the energy
density. Assuming the dilaton field is fixed for $\eta >
\eta_1$, since $H^2\propto \rho_A$, and $H=a'/a^2$, it is
straightforward to derive that $a\propto \eta^{1+\zeta}$. In this
case, with $\zeta>0$, the dynamics is driven by the axion
fluctuations. This is inconsistent since the gravitational
amplification of these fluctuations, in particular the spectral slope
of the axion, were calculated assuming that the fluctuations would
re-enter during a radiation dominated era, and with $\zeta>0$, the
expression for the scale factor shows that this is not the case.

This inconsistency reflects the breakdown of the perturbative approach
used to calculate the amplification of axion fluctuations.  In effect,
the calculation assumes that the quantum fluctuations can be treated
as a perturbation on a fixed classical background, whereas in the
present case, one should consider their back-reaction effect on the
background spacetime. Moreover, since the axion field is assumed not
to participate to the dynamics, its classical vev is zero, and the
energy density stored in axion is $\rho_A \sim (\partial\delta A)^2$,
where $\delta A$ represents the fluctuation in the axion field. One
should therefore include back-reaction up to second order in the
fluctuations, in order to derive the dynamics of the era between times
$\eta_1$ and $|\eta_s|$, and such an intricate calculation is well
beyond the scope of the present paper. Note, however, that the above
inconsistency does not arise for conformal times $\eta>|\eta_s|$,
since $n_j^{\rm\tiny DDI}>0$ for all $j$, and 
radiation domination should be a valid approximation.

We thus consider, as an alternative, that black holes form on all
scales comprised between $k_1$ and $k_s$. This is a possible outcome
of the above dynamics, as black holes generically form copiously when
relative overdensities of order unity re-enter the horizon.  At late
times $\eta > |\eta_s|$, the Universe will be dominated by those black
holes that have not evaporated yet. The lifetime of a black hole
$\tau_{\rm bh}\propto M_{\rm bh}^3\propto H_i^{-3}$, where $M_{\rm
bh}$ denotes the black hole mass, and corresponds to the mass within
the horizon at the time of formation at Hubble scale $H_i$. Therefore
the Universe will be dominated by those black holes that formed last,
i.e. on scale $k_s$, and will reheat with the evaporation of those
black holes. The Hubble scale $H(|\eta_s|)\sim H_1
|\eta_s/\eta_1|^{-3/2}$ if the era between times $\eta_1$ and
$|\eta_s|$ is matter dominated (black hole domination). Then $M_{\rm
bh}\sim 4\pi\mpl[H(|\eta_s|)/\mpl]^{-1}$ (which corresponds to the
mass within the horizon at that time) and the evaporation timescale of
those black holes reads $\tau_{\rm bh}\sim \mpl^{-1} (M_{\rm
bh}/\mpl)^3\sim (4\pi)^3 \mpl^{-1}[H(|\eta_s|)/\mpl]^{-3}$, so that
the black holes evaporate and reheat the Universe at a Hubble scale
$H_{\rm ev}\sim (4\pi)^{-3}H(|\eta_s|)[H(|\eta_s|)/\mpl]^2$.

Such reheating by black hole evaporation was envisaged in
Ref.~\cite{CLLW98} in the context of the PBB scenario. Note that it is
accompanied by entropy production, to the level of $\Delta s \sim
(\rho_{\rm BH}/\rho_r)^{3/4}$, where $\rho_{\rm BH}$ and $\rho_r$
denote the energy densities contained in black holes and in radiation
at the time of evaporation. This entropy production can be rewritten
as $\Delta s\sim
10^{19}\,g_1^{1/2}\Omega_r^{-3/4}[H(|\eta_s|)/10^7\,{\rm GeV}]^{-3/2}
(H_1/M_s)^{1/2}$, and the reheating temperature $T_{\rm RH}\sim
0.6\,{\rm GeV}\,[H(|\eta_s|)/10^7\,{\rm GeV}]^{3/2}$.  Even though
this entropy production may be sufficiently large to dilute the moduli
created gravitationally during the DDI and ``string'' phases, the same
moduli are also part of the Hawking radiation of the evaporating black
holes. The number density to entropy density ratio of moduli and
gravitinos present {\it after} evaporation in fact reads~\cite{L00}:
$Y_m\sim10^{-9} [H(|\eta_s|)/10^7\,{\rm GeV}]^{1/2}$, assuming
$g_\star\sim200$ at evaporation.  Clearly, black hole evaporation
eliminates one moduli problem, to reintroduce another, and further
entropy production is necessary. 

Consider then a modulus $\chi$ as introduced in Section IV.A, with
mass $m_\chi$. The amount of entropy produced by $\chi$ is given in
Eq.~(\ref{dS_chi}). However, if black holes have not evaporated by the
time $\chi$ would dominate (if black holes were absent), the r.h.s of
Eq.~(\ref{dS_chi}) should be multiplied by $(H_{\rm ev}/H_{\rm
dom})^{1/2}$, where $H_{\rm dom}\sim m_\chi (\chi_0/\mpl)^4$
corresponds to the Hubble scale at which $\chi$ would dominate in the
absence of black holes. Using $H_{\rm ev}\sim 10^6\,{\rm GeV}
[H(|\eta_s|)/2 \cdot10^{15}\,{\rm GeV})]^3$, and assuming $H_{\rm
ev}\la H_{\rm dom}$, the final moduli abundance is:

\beq
Y_m \sim 10^{-17} \,\gamma_\chi^{1/2}\,
\left(\frac{m_\chi}{10^6\,{\rm GeV}}\right)^{3/2}\,
\left(\frac{H(|\eta_s|)}{2 \cdot 10^{15}\,{\rm GeV}}\right)^{-1}
\,\Omega_r^{3/4}\,,
\eeq
hence, for $H_1\sim 10^{17}\,$GeV, provided the phase in which black
holes dominate does not last too long, i.e. $|\eta_s/\eta_1|\la
4\times10^3$, then $H(|\eta_s|)\ga 4\times10^{11}\,{\rm GeV}$ and
$Y_m\la10^{-13}$. 

To summarize, black hole reheating of the PBB scenario, as envisaged in
Ref.~\cite{CLLW98}, does not solve the moduli problem; it also
requires another source of entropy production, and the oscillations
and decay of the $\chi$ modulus would be sufficient, provided the
``string'' phase does not last more than $\sim8$ e--folds of the scale
factor. In this case, baryogenesis could also be implemented as in
Section IV.B.

\section{Discussions and conclusions}
\label{sec6}

We find that pre-big bang cosmological models inevitably face a severe
gravitino/moduli problem, as they predict a number density to entropy
density ratio of gravitationally produced moduli at the beginning of
the radiation era of the order of $Y_{m} \sim 0.3/g_\star$, where
$g_\star$ counts the number of degrees of freedom in the radiation
bath at that time. These models also predict a similar amount of
gravitinos, albeit somewhat smaller if gravitinos are not produced
gravitationally during dilaton driven inflation, yet far in excess
of the BBN bounds $Y_{m,3/2}\la 10^{-13}$. 

Late entropy production, to the level of $\Delta s\ga 10^5 - 10^{10}$
depending on the details of the transition between the pre-big bang
inflationary era and the radiation phase, is thus mandatory in the
scenarios we have investigated. For the simplest pre-big bang model
in which the transition is sudden, the amount of entropy needed is
$\Delta s\ga10^{10}$ and this is a strong requirement. However,
sufficient entropy can be produced by the domination and decay of the
zero-mode of a modulus field with mass $\sim 10^6\,$GeV, initially
displaced from the minimum of its potential by an amount
$M_s$. The Universe will start the FRW radiation 
era with a temperature  $T_{\rm RH} \sim 1\,{\rm GeV}$ (see Eq.~(\ref{4.4})). 
Moreover, the dilaton, which drives the pre-big bang dynamics,
could also play the role of this modulus, as several scenarios of
gaugino condensation predict a dilaton mass $\sim10^6\,$GeV, and since
it can be generically displaced from its present value at the end of the
pre-big bang inflation. Furthermore, this vast amount of entropy
produced helps set the Affleck-Dine mechanism of baryogenesis in a
natural framework, as it reduces efficiently the baryon asymmetry
created in the decay of the baryon number carrying flat
direction. Finally, it may also solve the usual monopole problem
associated with GUT symmetry breaking, although this depends
sensitively on the details of monopole formation at the GUT phase
transition.

We also examined variants of the pre-big bang model in which an
intermediate phase of dynamics motivated by physics at high curvature
takes place between the pre-big bang inflationary phase and the
radiation era. In the case of the so-called dual-dilaton intermediate
phase, one finds that the moduli/gravitino problem is still present,
and entropy production is still necessary. However the problem of
entropy production is relieved by the small fraction of energy density
contained in radiation at the beginning of the radiation era. In
effect, the energy density is generically contained in an axion field,
either the model-independent axion $A$ or internal axions $B_{ab}$
associated with the compactified components of $B_{\mu\nu}$. One can
show that several natural sources of entropy may alleviate or solve
the moduli/gravitino problem, notably the entropy produced in axion
reheating via scattering (provided the axion decay constant $F\la
10^{-3}\Mpl$), or that produced in oscillations and decay of the
zero-mode of the dominating axion, if the axion decay constant
$\sim10^{-3}\Mpl$ and its potential is generated by gaugino
condensation in a hidden sector at scale $\Lambda\sim\,{\rm
few}\times10^{13}\,$GeV. In some regions of parameter space (which
parametrizes the evolution of internal and external dimensions), the
amount of entropy needed to reduce the moduli/gravitino problem is
sufficiently small ($\Delta s\sim10^6$) that the domination and decay
of an Affleck-Dine condensate can produce both the entropy and the
baryon asymmetry of the Universe. 
In this case the hot big bang, which marks the beginning 
of the FRW radiation era, takes place 
at a temperature $T_{\rm RH} \sim 10^5\,{\rm GeV}$ (see Eq.~(\ref{4.10})).

In the case of the so-called ``string'' intermediate phase, one is at
present unable to specify the dynamics of the era that follows the
string phase (see Section VI). However, as we have argued, it is
likely that microscopic black holes would form copiously. Black hole
domination and decay produces entropy, which would dilute the
moduli/gravitinos produced during the dilaton-driven and string
phases, but moduli and gravitinos are also re-created in the Hawking
radiation of the evaporating black holes. Here again, therefore,
further entropy production is necessary, and the decay of a heavy
modulus can produce sufficient entropy.

At this stage we would like to comment on the implication of entropy
production on the various predictions of the pre-big bang models.
First of all the entropy production will not affect in any way the
axion seeds of large scale structure considered by Durrer et
al. \cite{DGSV98}. In effect, as long as the axion perturbations lie
outside the horizon, they are frozen, and do not suffer from the
microphysical processes inside the horizon, i.e. they are not diluted
by entropy production. More quantitatively, the density perturbation
$\delta\rho_A/\rho_c$ in the axion field relative to the total energy
density, can be written as a function of comoving wavenumber $k$ and
conformal time $\eta$, as~\cite{DGSV98}: $\delta\rho_A/\rho_c \sim
(k\eta)^2 (H_1/\Mpl)^2 (k/k_1)^{n_A^{\rm \tiny DDI}}$, for modes
outside of the horizon, i.e. $k<1/\eta$, and where $n_A^{\rm \tiny
DDI}$ denotes the axion spectral index of density
fluctuations. Clearly, when the mode re-enters the horizon,
i.e. $k=1/\eta$, for $n_A^{\rm \tiny DDI}\approx 0$ (scale invariant
spectrum), the density perturbation is independent of any entropy
production. Note however that a nearly flat spectrum of axion
fluctuations corresponds to a region of parameter space in which the
necessary amount of entropy release is large, of order $\Delta
s\sim10^{10}$.

Therefore, if perturbations of the $\chi$ field, which is the modulus
responsible for entropy production, do not carry power on large
scales, i.e. $n_\chi^{\rm\tiny DDI} > 0$, as would be the case if
$\chi$ were the dilaton for instance, the scenario envisaged by Durrer
et al. \cite{DGSV98} for the axion seeds remains unaffected. In this
framework, the pre-big bang predicts non-Gaussian isocurvature
perturbations with a well defined signature in the small angular scale
cosmic microwave background anisotropies. However, if the perturbations $\delta\chi$ carry power on
large scales, adiabatic perturbations would be produced on these
scales at $\chi$ decay, since it dominates the evolution at that
time. The study conducted in Ref.~\cite{BMUV98} seems to indicate that
the only fields in pre-big bang models that are liable to carry power
on large scales are the axions $A$ or $B_{ab}$. In this context, the
domination and decay of an axion, as considered in Section V.A., could
lead to a novel scenario of generation of density perturbations in
pre-big bang models. One should calculate carefully and examine the
exact shape of the spectrum of metric fluctuations and their
statistics, as it is known that axionic fluctuations are generally
damped on large scales due to the periodic nature of the
potential~\cite{KL87}. 

Scalars generically carry steep blue fluctuations spectra in
pre-big bang models, hence neither the modulus $\chi$ nor the
Affleck-Dine condensate $\Phi$ envisaged in Section IV are liable to
produce long wavelength fluctuations at their decay. This is in some
contrast to standard inflationary models, in which the decay of the
Affleck-Dine field produces isocurvature long wavelengths
fluctuations.

On similar grounds, one does not expect that the spectrum of
electromagnetic fields on the scale of the Galaxy should be diluted by
entropy production, as these fluctuations were outside of the horizon
at the time at which entropy was released. However, one expects that
the relic gravitational wave background will be at least partly
affected by the entropy release~\cite{BGV97,CLLW98}. Whether this
dilution affects the stochastic gravitational background in the range
of frequencies which the upcoming experiment, LIGO, Virgo and LISA,
are sensible deserves further investigation.

\acknowledgments We wish to thank Les Houches Summer School 1999 {\em
The Primordial Universe}, where this work was initiated. We would like to
thank A. Linde for interesting discussions and R. Brustein for 
useful comments.

AB thanks the D\'epartement d'Astrophysique Relativiste et de
Cosmologie in Observatoire de Paris-Meudon (France) for hospitality.
AB's research was supported by the Richard C. Tolman Fellowship and by
NSF Grant AST-9731698 and NASA Grant NAG5-6840.  The work of KAO was
supported in part by the Department of Energy under Grant No.\
DE-FG-02-94-ER-40823 at the University of Minnesota.

\end{document}